\documentclass[fleqn,usenatbib]{mnras}

\usepackage{newtxtext,newtxmath}
\usepackage[T1]{fontenc}

\usepackage{amsmath}
\usepackage{graphicx}
\usepackage{subcaption}
\usepackage{threeparttable}

\DeclareRobustCommand{\VAN}[3]{#2}
\let\VANthebibliography\thebibliography
\def\thebibliography{\DeclareRobustCommand{\VAN}[3]{##3}\VANthebibliography}

\title[Abell 272]{Understanding the Nature of the Ultra-Steep Spectrum Diffuse Radio Source in the Galaxy Cluster Abell 272}
\author[A. Whyley et al.]{
A. Whyley$^{1,2}$\thanks{Email: arthur.whyley@port.ac.uk},
S. W. Randall$^{3}$,
T. E. Clarke$^{4}$,
R. J. van Weeren$^{5}$,
K. Rajpurohit$^{3}$,
W. R. Forman$^{3}$,
\newauthor
A. C. Edge$^{6}$,
E. L. Blanton$^{7}$,
L. Lovisari$^{8,3}$,
and H. T. Intema$^{5}$
\\
\\
$^{1}$Institute of Cosmology and Gravitation, University of Portsmouth, Dennis Sciama Building, Portsmouth, PO1 3FX, UK\\
$^{2}$Department of Physics and Astronomy, University of Southampton, University Road, Southampton, SO17 1BJ, UK\\
$^{3}$Center for Astrophysics $|$ Harvard $\&$ Smithsonian, 60 Garden Street, Cambridge, MA 02138, USA\\
$^{4}$U.S. Naval Research Laboratory, 4555 Overlook Avenue SW, Washington, DC 20375, USA\\
$^{5}$Leiden Observatory, Leiden University, PO Box 9513, 2300 RA Leiden, The Netherlands\\
$^{6}$Centre for Extragalactic Astronomy, Department of Physics, Durham University, South Road, Durham, DH1 3LE, UK\\
$^{7}$Institute for Astrophysical Research and Department of Astronomy, Boston University,  Boston, MA 02215, USA\\
$^{8}$INAF, Istituto di Astrofisica Spaziale e Fisica Cosmica di Milano, via A. Corti 12, 20133 Milano, Italy}

\pubyear{2024}

\begin{document}
\maketitle
\label{firstpage}
\pagerange{\pageref{firstpage}--\pageref{lastpage}}

\begin{abstract}

Ultra-steep spectrum (USS) radio sources with complex filamentary morphologies are a poorly understood subclass of diffuse radio source found in galaxy clusters. They are characterised by power law spectra with spectral indices less than -1.5, and are typically located in merging clusters. We present X-ray and radio observations of the galaxy cluster A272, containing a USS diffuse radio source. The system is an ongoing major cluster merger with an extended region of bright X-ray emission south of the core. Surface brightness analysis yields a $3\sigma$ detection of a merger shock front in this region. We obtain shock Mach numbers $M_\rho = 1.20 \pm 0.09$ and $M_T = 1.7 \pm 0.3$ from the density and temperature jumps, respectively. Optical data reveals that the system is a merger between a northern cool core cluster and a southern non-cool core cluster. We find that the USS source, with spectral index $\alpha^{\text{74 MHz}}_{\text{1.4 GHz}} = -1.9 \pm 0.1$, is located in the bright southern region. Radio observations show that the source has a double-lobed structure with complex filaments, and is centred on the brightest cluster galaxy of the southern subcluster. We provide two suggestions for the origin of this source; the first posits the source as an AGN relic that has been re-energised by the passing of a merger shock front, while the second interprets the complex structure as the result of two overlapping AGN radio outbursts. We also present constraints on the inverse Compton emission at the location of the source. 

\end{abstract}


\section{Introduction}

Galaxy clusters are the largest gravitationally bound objects in the universe. They contain three main components; galaxies, dark matter, and the intracluster medium (ICM). %The number of galaxies in a cluster can range from around a hundred to more than a thousand. Dark matter dominates the mass of clusters and is therefore responsible for holding them together through gravitational attraction. Its mass typically accounts for 80 percent of the total cluster mass \citep{Bykov2015}. The space between the galaxies is filled with the ICM.
The ICM has an extremely low density and is weakly magnetised. Typical ICM temperatures range from 1-10 keV. %(1 keV $= 1.16 \times 10^7$ K).
Emission from the ICM is dominated by thermal bremsstrahlung and line emission, and emits most strongly in the X-ray band \citep[0.1-10 keV; e.g.][]{Mohr1999}.
Galaxy clusters primarily grow by merging with other clusters. %This occurs when two clusters are drawn together by gravity and eventually become one larger cluster. 
Mergers are the most energetic events in the universe since the Big Bang, and take place over very long (Gyr) timescales. During mergers, large amounts of kinetic energy are dissipated into the ICM via shocks and turbulence as a result of the colliding ICM plasma \citep[e.g.][]{Markevitch1999}. %This heats the ICM and causes further X-ray emission \citep[e.g.][]{Markevitch2007}. %It is extremely rare for galaxies to collide in a merger (due to the distances between galaxies being much larger than the sizes of the galaxies), and dark matter is almost completely collisionless \citep{Randall2008,Garrett2010,Molnar2016}. 

There are multiple mechanisms that can give rise to diffuse radio emission in clusters that is not directly associated with radio-emitting galaxies \citep{Feretti2012,van2019diffuse}. %Most observed radio emission of this type occurs in merging clusters. This has been observed in over 100 clusters \citep{van2019diffuse}. 
Diffuse radio sources in clusters can be divided into four main categories: radio halos, radio relics, radio phoenices and active galactic nuclei (AGN) relics \citep{2004rcfg.proc..335K}.

Radio halos are located in the centre of a cluster and have a regular morphology, with their emission typically following the thermal distribution of the ICM \citep[e.g.][]{govoni2001radio, Kassim2001}. Typical halos are hundreds of kpc in size \citep[e.g.][]{giacintucci2013discovery} and their radio emission is not observed to be polarised \citep[e.g.][]{Shimwell2014}. Halos are commonly found in massive clusters and are currently thought to be caused by the acceleration of electrons via merger turbulence, which results in the observed synchrotron radio emission \citep[e.g.][]{Brunetti2001,2014IJMPD..2330007B}. 

The cool cores of relaxed clusters can display evidence of faint and extended diffuse emission surrounding their central AGN, referred to as a radio mini-halo \citep[see e.g.][and references therein]{2019ApJ...880...70G, 2020MNRAS.499.2934R}. Mini-halos are thought to be powered by turbulent acceleration and are observed to be confined by X-ray cold fronts in the ICM. They are typically located in the central cool core region with sizes of 50 kpc to $0.2R_{500}$ \citep{2017ApJ...841...71G} although some systems show larger-scale diffuse emission surrounding them \citep[e.g.][and references therein]{10.1093/mnras/stac672}. %Recently, Giacintucci et al (2024) showed that the larger scale diffuse radio emission in both Abell 3444 and MS 1455.0+2232 is confined by newly-identified large-scale sloshing cold fronts, similar to more compact systems confined by central cold fronts.

Radio relics are another type of diffuse radio source that are observed in clusters. Unlike halos, they are located in the outskirts of a cluster, have elongated shapes \citep[that typically span up to a few Mpc; e.g.][]{de2014new}, and emit highly polarised emission \citep[e.g.][]{Slee2001,2011A&A...528A..38V}. They are thought to be produced by the acceleration of cosmic rays by merger driven shocks and are typically found at shock fronts \citep[e.g.][]{Giacintucci2008}.

Radio phoenices are typically smaller than halos or relics, with sizes ranging from tens to hundreds of kpc, and their location and morphology can vary greatly. Presently, it is thought that they are caused by the adiabatic compression of the ICM plasma as result of merger shocks \citep[e.g.][]{ensslin2002formation,Enblin2001, Gasperin2015}.

Radio emission observed in clusters can also be due to AGN; outbursts from AGN disperse radio plasma into the surrounding ICM. This plasma can then appear to be a diffuse source that is ``detached" from the AGN. These are classified as AGN relics \citep{2004rcfg.proc..335K}. 

Ultra-steep-spectrum (USS) sources are characterised by radio power law spectra with spectral indices $\alpha < -1.5$\footnote{Here, and throughout the paper, $\alpha$ is defined by the relationship $F_\nu \propto \nu^\alpha$.} between 74 MHz and 1.4 GHz \citep[e.g.][]{Slee2001,Feretti2012}, that are often curved at higher frequencies. They are a relatively poorly understood type of diffuse radio source, and have been observed in several clusters. Examples include A980 \citep{salunkhe2022deciphering}, A2877 \citep{hodgson2021ultra}, RX J1720. 1+ 2638 \citep{biava2021ultra} and A521 \citep{santra2023deep}, as well as those discussed in \cite{mandal2020revealing}. They are typically observed in merging clusters, suggesting that mergers are responsible for their formation. USS sources can be found in cluster cores, but here we only consider ones found outside cores, to avoid confusion with radio activity related to the central AGN. In particular, we focus on diffuse sources with complex morphologies, that lack a clear host galaxy. Possible origins of USS emission include: (1) ``weak" radio relics that occur due to minor mergers, (2) AGN relics, or (3) radio phoenices \citep{2004rcfg.proc..335K}. It is currently unknown whether just one of these mechanisms is responsible for all observed USS sources, or if all three (or only two) can give rise to USS emission.

It was suggested by \cite{Randall2016} that some of the different types of diffuse radio sources in merging galaxy clusters may be part of a single ``lifecycle", after they found evidence for the existence of a radio relic and a radio phoenix in the same cluster. They propose that AGN relics could become radio phoenices after the relic plasma is re-energised by adiabatic compression. Over time, a phoenix could break apart, with its non-thermal electron population spreading throughout a larger region of the ICM. This population could then be reaccelerated by a shock, giving rise to a radio relic. 

The study of USS sources provides an opportunity to find links between the properties of the observed radio source and the properties of its parent cluster. Currently, there are only a few systems in which USS sources have been shown to have connections to merger shocks \citep[e.g.][]{Cohen2011,Macario2011}. By finding these connections in more clusters, the current theories of ICM dynamics in mergers (and how they lead to USS emission) can be confirmed or challenged. This is significant because it will further the understanding of particle interactions in a hot%, high-$\beta$
, weakly collisional plasma - something that cannot be easily studied in the laboratory or in other astronomical systems.

In this paper, we present XMM-Netwon X-ray observations and VLA, LOFAR and GMRT radio observations of the galaxy cluster Abell 272. The cluster is at redshift $z$ = 0.0877 \citep{Struble1991} and hosts a non-central USS diffuse radio source, with spectral index $\alpha^{\text{74 MHz}}_{\text{1.4 GHz}} = -1.9 \pm 0.1$ (calculated with archival VLA data). We show an optical image of the cluster in Figure \ref{OpticalIm}. A272 was selected for follow-up X-ray and radio observations to further study the system, in a similar manner to the analyses presented in \cite{Clarke_2013} and \cite{Randall2016}. This paper aims to determine the dynamical state of Abell 272, and use the X-ray and radio observations to investigate the nature and origin of the USS radio emission. We use a cosmology with $\Omega_0 = 0.3,$ $\Omega_{\Lambda} = 0.7$ and $H_0 = 70$ km s$^{-1}$ Mpc$^{-1}$, such that $1'$ corresponds to a distance of 112 kpc at the cluster redshift. All uncertainties represent the 1$\sigma$ confidence interval, unless otherwise stated. 

\section{Data Analysis}

\subsection{X-Ray Image Creation}

Abell 272 was observed by XMM-Newton for 18.6 ks on 13th August 2018 (ObsID 0821690301). The data were processed using XMM SAS version 20.0.0\footnote{\url{https://www.cosmos.esa.int/web/xmm-newton/sas}} and the XMM ESAS package version 5.9\footnote{\url{https://www.cosmos.esa.int/web/xmm-newton/xmm-esas}}. Data reduction was performed using the standard ESAS pipeline of \cite{cookbook}. First, we filtered the raw events files using the tasks \verb|emchain| and \verb|mos-filter| for the MOS1 and MOS2 data, and the tasks \verb|epchain| and \verb|pn-filter| for the PN data. This removed any parts of the data that were contaminated due to soft proton flaring. Point sources were then identified and removed with the task \verb|cheese|. This automatically detected sources by searching for bright regions in the image (with fluxes $> 10^{-14}$ ergs cm$^{-2}$ s$^{-1}$) that were separated from any other bright regions by at least $40''$. Following this, we checked the mask images produced by the task and manually added any missed point sources to the source list file. The task \verb|make_mask| was then used to add these additional sources to the masks and remove them from the counts images. 

The next step was to create model images of the quiescent particle background (QPB) and soft proton background, which are both background components present in XMM-Newton data \citep{ParticleBkg}. The QPB is the result of high energy particles interacting with the detectors and the surrounding structure. The soft proton background refers to cosmic ray protons that have the right energies for them to be focused by the telescope mirror, and counted by the detectors as if they are X-rays. We used the tasks \verb|mos-spectra| and  \verb|pn-spectra|, for MOS and PN respectively, to extract the background spectrum from an annulus outside the central region of the cluster for each detector. The spectra were fitted simultaneously to determine the parameters of the soft proton background. (See Section 2.2 for a detailed description of the spectral analysis process.) The soft proton parameters were used in the ESAS task \verb|proton| to create a model image of the soft proton background for each detector. We also created model QPB images for each detector, by using the tasks \verb|mos_back|, for MOS, and \verb|pn_back|, for PN. Following this, the task \verb|comb| was used to combine the counts images for each detector (in the 0.4 - 7.0 keV band) into a single total counts image, and then subtract the soft proton background and QPB images. Finally, the resultant image was then corrected for its exposure time and binned to $5''$ pixels in order to create a count rate image. This was done using the task \verb|bin_image|. We show the count rate image in Figure \ref{X-ray}.

\subsection{X-Ray Spectral Analysis}

All X-ray spectral analysis was performed using XSPEC version 12.12.1 \citep{XSpec}. \cite{GrSa} abundances ratios and AtomDB version 3.0.9 \citep{AtomDB} were used for all fits. In all cases, spectra were extracted from each of the three detector images (MOS1, MOS2 and PN) and fitted simultaneously in the 0.4 - 7.0 keV band with model parameters linked between them, unless stated otherwise. In the cases where a parameter was linked, we tested allowing it to vary independently for each detector and confirmed that the three values were consistent at the 1$\sigma$ level. The thermal bremsstrahlung and line emission from the ICM were modeled with an absorbed APEC model, with a fixed Galactic absorption $n_H = 0.049$ x $10^{22}$ cm$^{-2}$ \citep{nH}. The $\chi^2_\nu$ statistic was used for all spectral fitting.

Various other model components were used to account for the background emission. Their parameters were initially determined by fitting the spectra extracted from an annulus outside the central region of the cluster, with inner and outer radii of $7.3'$ and $10.4'$ respectively. This region was chosen so that the ICM emission in each spectrum was negligible. The background consisted of three parts that needed to be modeled - the X-ray sky background, the instrumental lines present in XMM-Newton spectra and the soft proton background.

\begin{table*}
	\centering
	\caption{Best fit normalisation values for the three X-ray sky background model components.}
	\label{tab:skybkg}
	\begin{tabular}{ccc} % four columns, alignment for each
		\hline
		Local Hot Bubble & Galactic Halo & Cosmic X-ray Background\\
        (cm$^{-5}$ arcmin$^{-2}$) & (cm$^{-5}$ arcmin$^{-2}$) & (photons keV$^{-1}$ cm$^{-2}$ s$^{-1}$ arcmin$^{-2}$)\\
        \hline
		  (4.9 $\pm$ $0.1)\times10^{-7}$ & (1.04 $\pm$ $0.06)\times10^{-6}$ & (9.5 $\pm$ $0.9)\times10^{-7}$\\
		\hline
	\end{tabular}
\end{table*}

\begin{table*}
	\centering
	\caption{Best fit parameters for the soft proton background power law model, where $\alpha_1$ is the power law index before the break and $\alpha_2$ is the index after the break. All paramters listed were allowed to vary, apart from $\alpha_2$ in PN.}
	\label{tab:sp}
	\begin{tabular}{ccccc} % four columns, alignment for each
		\hline
		Detector & $\alpha_1$ & $\alpha_2$ & Break Energy (keV) & Normalisation (10$^{-4}$ arcmin$^{-2}$)\\
        \hline
        MOS1 & 0.53 $\pm$ 0.05 & 0.22 $\pm$ 0.06 & 3.0$^{+0.5}_{-0.2}$ & 3.0 $\pm$ 0.2\\
        \hline
		  MOS2 & 0.56 $\pm$ 0.04 & 0.17 $\pm$ 0.06 & 3.5$^{+0.2}_{-0.3}$ & 2.7 $\pm$ 0.1\\
		\hline
        PN & 0.4 $\pm$ 0.4 & 2.5 & 0.98$^{+0.05}_{-0.05}$ & 6 $\pm$ 1\\
		\hline
	\end{tabular}
\end{table*}

The following model components were used for the X-ray sky background: an unabsorbed APEC model for the Local Hot Bubble; an absorbed APEC model for emission from the Galactic halo; and an absorbed power law model for the cosmic X-ray background. To help constrain their parameters, we included the spectrum from the ROSAT All Sky Survey \citep{RASS} at the location of Abell 272 (extracted from an annulus between 0.5 and 1.0 degrees) which was obtained using the HEASARC X-Ray Background Tool \citep{RASStool}. This was fitted alongside the three XMM-Newton spectra, with its sky background parameters linked (apart from the normalisation of the power law, because of the different point source detection limits between XMM-Newton and ROSAT). The Local Hot Bubble and Galactic halo components used fixed temperatures $kT = 0.1$ and $kT = 0.21$ respectively, and both assumed solar abundance. The cosmic X-ray background power law had a fixed power law index of 1.45. These constants were all recommended values from the ESAS cookbook \citep{cookbook}. Table \ref{tab:skybkg} lists the best fit normalisations of the three sky background components.

The instrumental lines were modelled using Gaussians, with one Gaussian for each line centred on its energy. Our model differed between the two MOS spectra and the PN because of the different lines present in their spectra. In the 0.4 - 7.0 keV band, MOS spectra contain Al K$\alpha$, Si K$\alpha$, Cr K$\alpha$, Mn K$\alpha$ and Fe K$\alpha$ lines, whereas PN spectra contain Al K$\alpha$, Cr K$\alpha$ and Fe K$\alpha$ lines. For cases where lines were undetected, they were excluded from the model to reduce the number of fitted parameters. In addition, the normalisations for each instrument were not linked due to variations between detector responses.

%In addition, the normalisations for each detector were not linked, even though they should theoretically be consistent with one another for a given line. This choice was made due to the known cross-calibration issues between the three instruments \citep[e.g.][]{2009A&A...496..879M}.

The soft proton background was modeled with a broken power law. The parameters were not linked between detectors because of their different levels of soft proton contamination. The best fit parameters are listed in Table \ref{tab:sp}. In the case of the PN spectrum, the parameters were difficult to constrain due to the low signal to noise ratio, particularly at higher energies. The best fit value for the spectral index after the break was greater than 2.5, which was unreasonably large. We therefore chose to freeze its value at 2.5 for all further analysis, as recommended by the ESAS cookbook \citep{cookbook}. We note that this change had a negligible effect on the goodness of fit.

The background parameters were allowed to vary in all other spectral fits (for spectra extracted from both the source region and annular background region), but were always consistent with the values shown here within the 1$\sigma$ error range. This choice allowed for the uncertainties in the background parameters to be included in the overall errors on source parameters.

\subsection{Radio Observations}

\begin{figure*}
\centering
\includegraphics[width=0.9\textwidth]{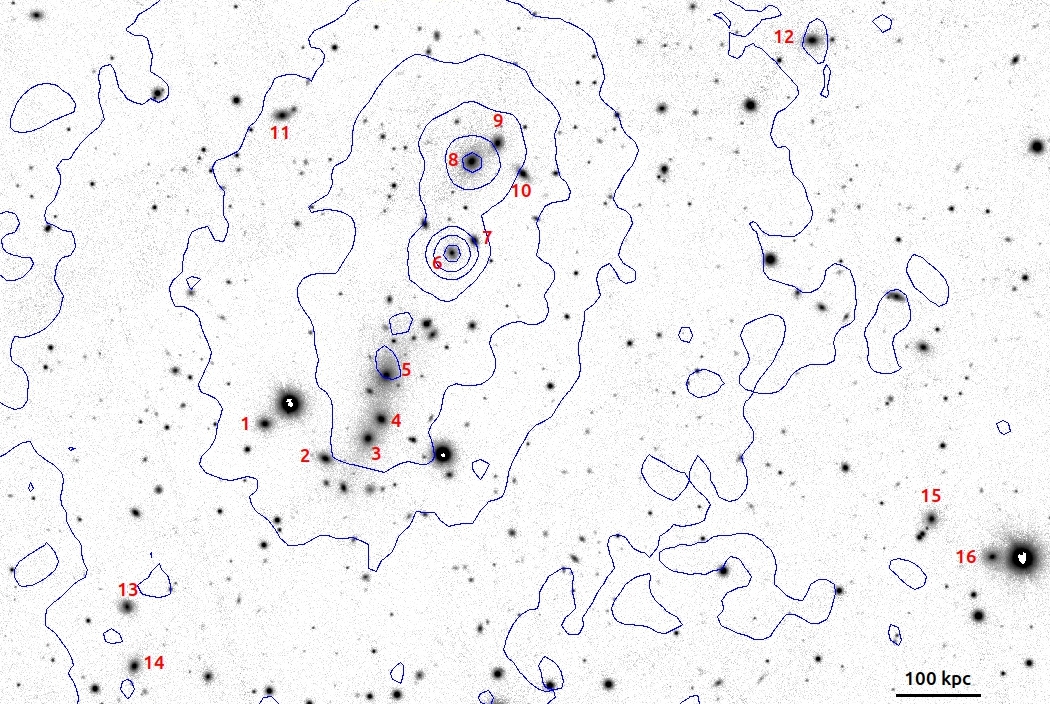}
\caption{\label{OpticalIm} Pan-STARRS \textit{I} band image of Abell 272 with XMM-Newton X-ray contours for the 0.4-7.0 keV band. 16 cluster members are numbered following \citet{sargent1972cluster}.}
\end{figure*}

%Here we outline the data that we utilise in our analysis of the radio emission in Abell 272. We use data from the 74 MHz Very Large Array Low-frequency Sky Survey redux \citep[VLSSr,][]{VLSS}. This includes an image as well as its associated flux measurement for the USS source. To provide a higher resolution view of the cluster, we use a Low Frequency Array \citep[LOFAR,][]{LOFAR} image. This has a frequency of 144 MHz. In addition, we also use a 325 MHz image from the Giant Metrewave Radio Telescope \citep[GMRT,][]{GMRT} as well as an existing GMRT flux value at 1.4 GHz.%

In this section we report both targeted observations of the radio emission toward Abell 272 as well as archival survey observations.

\subsubsection{Targeted GMRT Observations}

Radio continuum observations of A272 with the GMRT (Giant Metrewave Radio Telescope) at 325 MHz were performed on Jan 15, 2010. The cluster was observed for a total of 6 hrs, with 3C\,48 and 3C\,286 serving as the primary calibrators and 0137+331 as a secondary one. The data were reduced with the NRAO Astronomical Image Processing System (AIPS) package\footnote{\url{http://www.aips.nrao.edu}} \citep{2003ASSL..285..109G}.  The reduction consisted of a manual inspection for radio frequency interference (RFI) which was subsequently flagged. The bandpass and flux-scale were set using the primary calibrator sources 3C\,48, and 3C\,286 taking the \cite{perley1999vla} extension to the \cite{1977A&A....61...99B} scale. Gain solutions from the secondary calibrator source 0137+331 were then transferred to the target. For the target field, we first phase-calibrated against a skymodel created from the NVSS survey \citep{condon1998}. We then proceeded with several cycles of phase-only self-calibration and two last rounds of amplitude and phase self-calibration. Imaging was done using 37 facets to deal with the non-coplanar baselines \citep{1989ASPC....6.....P,1992A&A...261..353C}. We used a \cite{1995PhDT.......238B} weighting scheme with a robust value of 0.0 and the final image was corrected for the primary beam attenuation. For more details on the data reduction we refer the reader to \cite{2011A&A...527A.114V}.

\subsubsection{Archival Radio Surveys}

The VLA Sky Survey redux \citep[VLSSr;][]{lane2014} is an updated processing of the 74 MHz observations from the Very Large Array Sky Survey \citep[VLSS;][]{cohen2007}. The survey covers the entire northern sky (above dec -30) with a resolution of $75''$ and a sensitivity of 100 mJy/bm. %Overlaying the low resolution VLSSr radio image on higher resolution GMRT observations of A272 (described below), we confirmed that the VLSSr radio emission is spatially coincident with the filamentary radio structure shown in Figure \ref{LOFAR}. 
We extracted the radio emission for A272 from the VLSSr catalog, which has been corrected for clean bias and includes the measurement uncertainty and the flux density scale uncertainty. %The total flux of A272's filamentary emission is listed as 2.31 $\pm$ 0.45 Jy.

The Tata Institute of Fundamental Research (TIFR) Giant Metrewave Radio Telescope (GMRT) Sky Survey Alternative Data Release \cite[TGSS ADR;][]{intema2017} is a 150 MHz survey covering $\sim$ 90\% of the sky to a resolution of 25$^{\prime\prime}$. We extracted an image of the area around A272 and measure a local noise of 2.4 mJy/bm. Using this image, we measured the total flux of any relevant sources. Following \cite{intema2017}, a calibration uncertainty of 10\% was applied to these measurements.

The Low Frequency Array Two-meter Sky Survey \citep[LoTSS;][]{shimwell2019, shimwell2022} is an on-going northern sky survey conducted between 120 and 168 MHz with a spatial resolution of $5''$. A272 falls within the area of the second LoTSS data release that covers 27\% of the sky with a point-source sensitivity of 83 $\mu$Jy/bm. We extracted the full resolution LoTSS image of A272 and measured the total radio flux density of the USS source. %of the radio filaments to be 783.1 $\pm$ 1.3 mJy. 
Following \citet{shimwell2022}, a calibration uncertainty of 10\% was applied to these data.

The NRAO VLA Sky Survey \cite[NVSS;][]{condon1998} observed the entire northern sky above $\delta >$ -40 degrees at 1.4 GHz. The survey resolution is $45''$ with a typical noise of $\sim$0.45 mJy/bm. We identified the source of USS radio emission in the image and measured its flux density. %The radio emission from A272 appears as two regions of emission overlaid on the radio filaments. This is confirmed by the NVSS catalog which records two separate sources associated with A272 with a total flux of 7.1 $\pm$ 0.64 my.
Following \cite{condon1998} a calibration uncertainty of 3\% was applied to the measured values.

\subsection{Optical Observations}

We used optical data from the Sloan Digital Sky Survey \citep[SDSS,][]{SDSS} and Pan-STARRS \citep{chambers2019panstarrs1} to aid in our interpretation of the X-ray and radio results. The Pan-STARRS image of Abell 272 is shown in Figure \ref{OpticalIm}. The numbered galaxies in the image will be referenced throughout the paper using the notation G$\alpha$, where $\alpha$ is the galaxy number. Spectroscopic redshifts for 14 cluster member galaxies (1-6 and 8-15 in Figure \ref{OpticalIm}) are cited from \cite{sargent1972cluster}. 

\section{The Intracluster Medium}

\subsection{Global Properties}

The radius in which the mean density is 500 times greater than the critical density of the universe at the cluster redshift ($r_{500}$), was estimated using the $M_{500}-T_x$ (mass - X-ray temperature) scaling relation of \cite{r500},
\begin{equation}
    M_{500}=M_0(T/\text{5 keV})^\alpha E(z)^{-1}.
	\label{eq:Vikh}
\end{equation}

$M_{500}$ is the total mass within $r_{500}$, with $M_0 = 4.10\times10^{14} M_\odot$ and $\alpha = 1.5$. $E(z)$ is given by
\begin{equation}
    E(z) = H(z) / H_0
	\label{eq:Ez}
\end{equation}

for the cluster redshift. An iterative process was used; first, an estimate for $r_{500}$ was made and the spectrum was extracted within the range 0.15 $r_{500} < r < r_{500}$ (excluding emission at the location of G6 - see Section 3.2). We then fitted the spectrum to determine the average temperature within the region, which was used in the $M_{500}-T_x$ relation to calculate $M_{500}$. Next, the value of $r_{500}$ corresponding to this $M_{500}$ value was calculated using the definition of $r_{500}$, 
\begin{equation}
    r_{500}=\left(\frac{M_{500}}{500\times \frac{4}{3}\rho_c}\right)^\frac{1}{3}.
	\label{eq:r500}
\end{equation}

Here, $\rho_c$ is the critical density of the universe at the cluster redshift.
If this new $r_{500}$ was not consistent with the initial estimate, the process was then repeated, using the new value as the estimate, until they converged at the $3\sigma$ level. 

This method gave $r_{500}$ = 770 $\pm$ 20 kpc, $M_{500}$ = 1.4 $\pm$ 0.1$\times 10^{14} M_{\odot}$ and $T_{500}$ = 3.13 $\pm$ 0.06 keV (where $T_{500}$ is defined as the temperature of the region within 0.15 $r_{500} < r < r_{500}$). The cluster abundance was found to be $0.28^{+0.03}_{-0.02}$ $Z_\odot$.

\subsection{The X-Ray Image}

\begin{figure}
\centering
\includegraphics[width=0.45\textwidth]{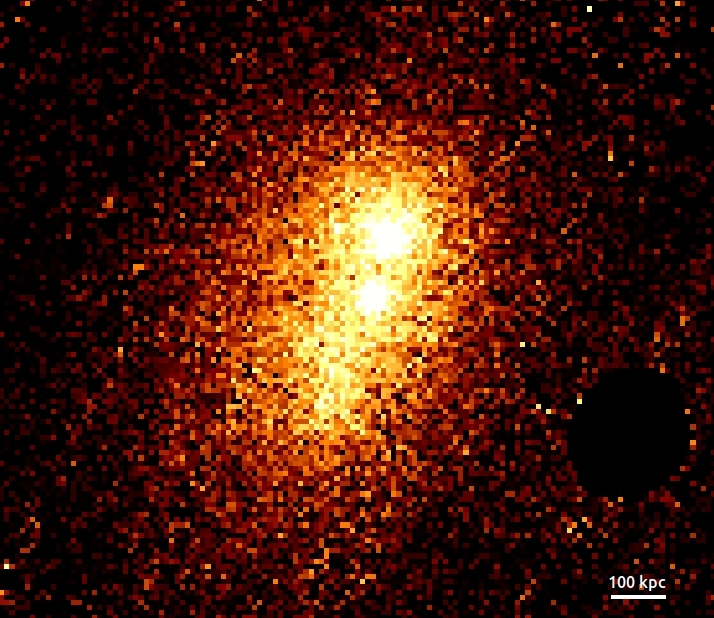}
\caption{\label{X-ray} Exposure corrected X-ray image of Abell 272 in the 0.4-7.0 keV band with the background subtracted and point sources removed. The ICM is elongated from N to S. The northern peak is the cluster core, and the southern peak is an AGN.}
\end{figure}

The X-ray image of Abell 272 is shown in Figure \ref{X-ray}, with all background and foreground point sources removed. The ICM is shown to be elongated south of the core, and has two prominent surface brightness peaks. The clear elongation indicates that the system is disturbed, and likely represents an ongoing major cluster merger. 

The image shows two bright peaks which lie along the axis of elongation, with the northern and southern peaks being centred on G8 and G6 from Figure \ref{OpticalIm}, respectively. The southern peak has X-ray contours consistent with an unresolved source (such as an AGN), while the northern peak has contours more consistent with extended emission (see Figure \ref{OpticalIm}). The spectra for both were extracted and fitted in XSPEC to constrain their nature. For each peak, we fitted the spectra first with an absorbed APEC model and then with an absorbed power law, to determine whether they were cluster cores or AGN (since AGN are characterised by power law spectra \citep{Ishibashi2010}, while ICM emission is thermal). The northern peak had $\chi^2_\nu = 1.264$ ($\nu = 540$) when fitted with an APEC model and $\chi^2_\nu = 1.446$ ($\nu = 537$) when fitted with an absorbed power law. We therefore concluded that it is the cool core of the system with kT = 1.63 $\pm$ 0.08 keV. This corresponds to G8, which is the brightest cluster galaxy (BCG). The southern peak had $\chi^2_\nu = 1.260$ ($\nu = 540$) when fitted with an APEC model and $\chi^2_\nu = 1.160$ ($\nu = 537$) when fitted with an absorbed power law. Since the power law better fitted the spectra, with a spectral index of 1.97 $\pm$ 0.05, we concluded that it was most likely an AGN. As this AGN is likely in G6, it is located within the cluster, and is therefore a potential second core of the system. However, because the emission is not consistent with thermal emission, the AGN was treated as a point source and excluded for any further analysis of the ICM.  

\subsection{Surface Brightness}

\begin{figure}
\centering
\includegraphics[width=0.45\textwidth]{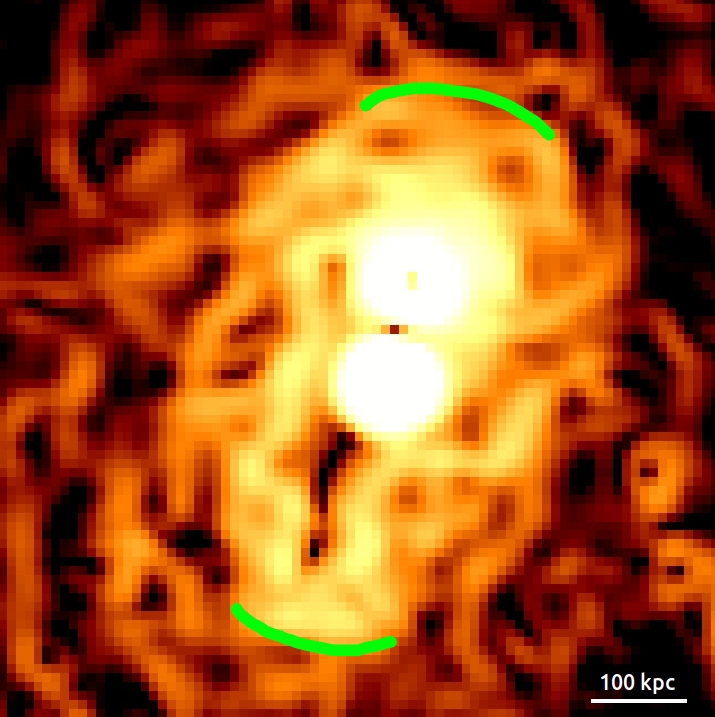}
\caption{\label{GGM} GGM filtered image of Abell 272, using a $\sigma$ = 5’’ Gaussian. Two potential shock fronts are highlighted with green lines.}
\end{figure}

\begin{figure}
    \begin{subfigure}[b]{0.2\textwidth}
    \centering
    \includegraphics[width=1.1\textwidth]{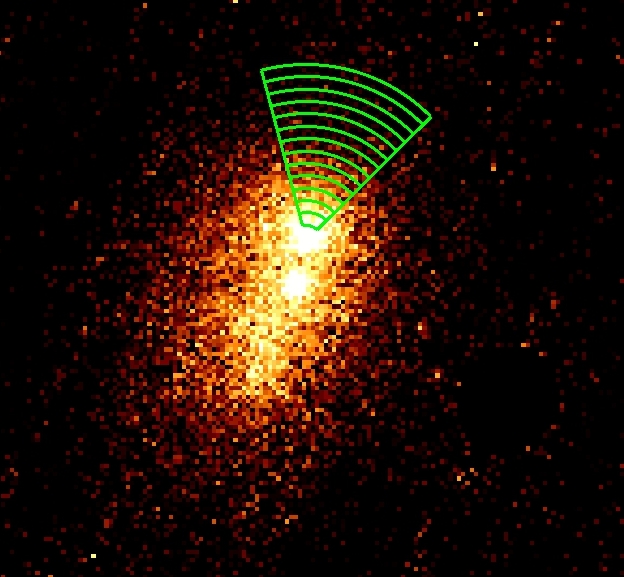}
    \end{subfigure}
    \hspace{4mm}
    \begin{subfigure}[b]{0.2\textwidth}
    \centering
    \includegraphics[width=1.1\textwidth]{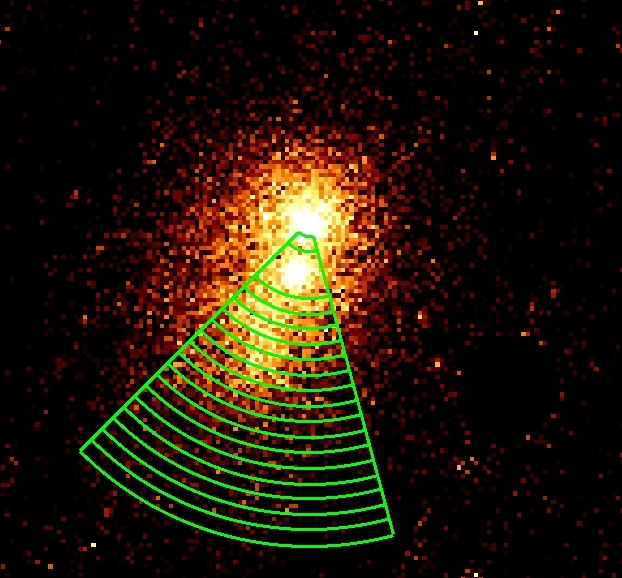}
    \end{subfigure}
\caption{\label{SBbins} Left: X-ray image of Abell 272 with the bins used for the northern surface brightness profile shown in green. Right: X-ray image of Abell 272 with the bins used for the southern surface brightness profile shown in green. The resulting profiles are shown in Figure \ref{SB}.}
\end{figure}

\begin{figure*}
\centering
    \begin{subfigure}[b]{0.45\textwidth}
    \centering
    \includegraphics[width=1.1\textwidth]{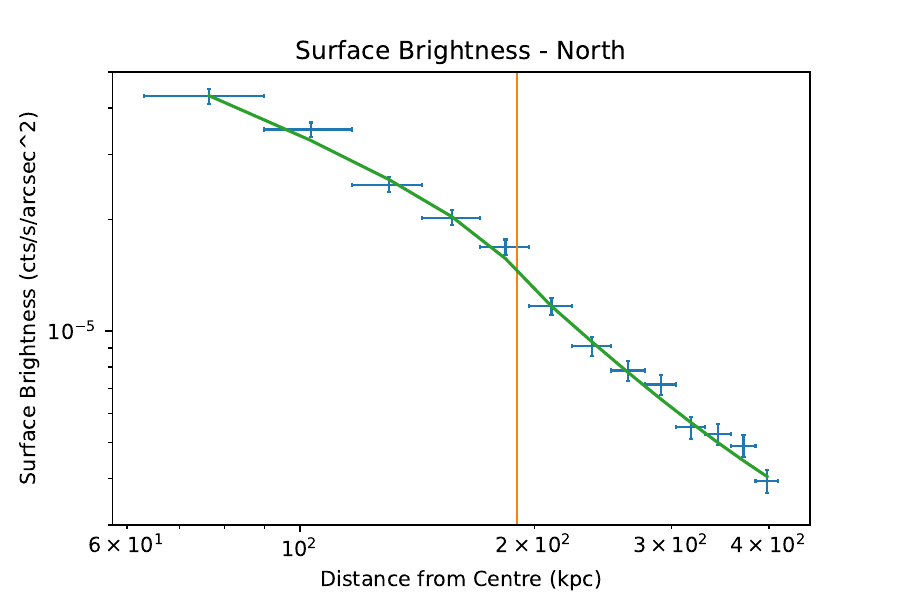}
    \end{subfigure}
    \hspace{4mm}
    \begin{subfigure}[b]{0.45\textwidth}
    \centering
    \includegraphics[width=1.1\textwidth]{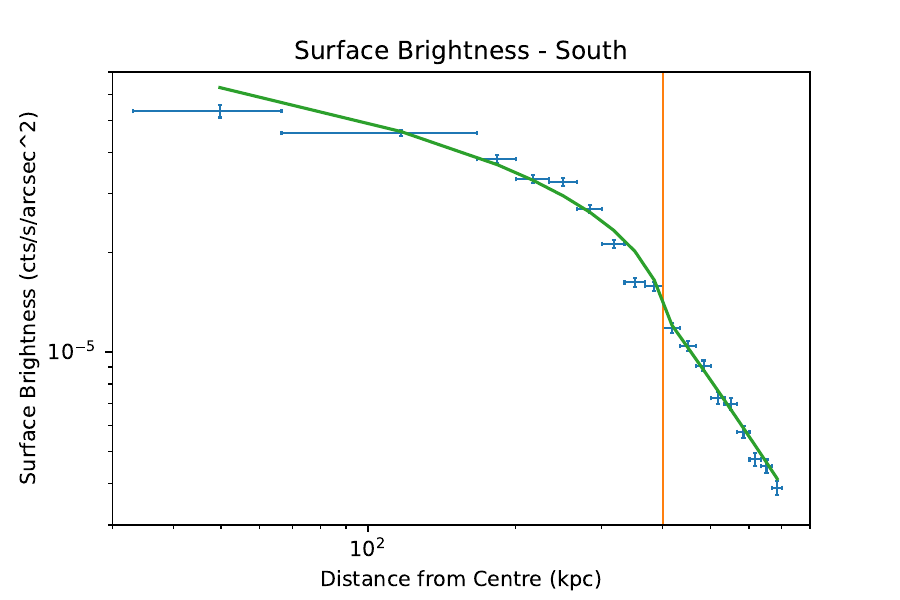}
    \end{subfigure}
\caption{\label{SB} Left: Surface brightness profile to the north of the cluster core. The fitted broken power law is shown in green and the vertical orange line marks the location of the potential shock front ($r = 190$ kpc). Right: Surface brightness profile to the south of the core. The fitted broken power law is shown in green and the vertical orange line marks the location of the potential shock front ($r = 400$ kpc)}
\end{figure*}

Major cluster mergers are expected to drive leading bow shocks in the ICM. Assuming that the merger axis is along the southern extension, we expect to find shocks somewhere on this axis. To search for potential shock fronts, we created a Gaussian gradient magnitude (GGM) filtered image of the cluster. %This process works by first creating the convolution of the total count rate image with a Gaussian. The gradient is then calculated at every point in the convolution. Finally, the output image is created by setting the value of each pixel to the magnitude of the gradient at that point. 
This process can be used to reveal substructures in the cluster by highlighting areas in which the surface brightness is changing rapidly with position \citep{GGM1,GGM2}. The GGM filtered image, using a $\sigma$ = 5$''$ Gaussian, is shown in Figure \ref{GGM}. Two bright arcs that could be potential shock fronts, one in the north and one in the south, are traced with green lines. 

To determine if these were shock fronts, we created surface brightness profiles to look for a significant variation in surface brightness across each potential front. Figure \ref{SBbins} shows the bins used%, with the bins for the northern profile in the left-hand panel and the bins for the southern profile in the right-hand panel.
. We chose the bins such that they extended radially outward from the core along the putative merger axis in both directions, and contained the bright extended region to the south. The surface brightness in a single bin was calculated by taking the total counts in the bin and subtracting the total counts from all background sources. This was done for each detector and the resultant counts from each were summed. Finally, we divided by the exposure time and bin area to get the surface brightness of the bin. Uncertainties were estimated as $\sqrt{N}$ whenever a counts value was taken from an image, and were propagated through. The two surface brightness profiles are shown in Figure \ref{SB}. %with the northern profile in the left-hand panel and the southern profile in the right-hand panel.
An orange line on each shows the position of the potential shock fronts. We fitted both profiles with a broken power law model projected along the merger axis. The model used was of the form
\begin{equation}
    I(r) = I_0\int F(\omega)^2dl
	\label{eq:I(r)}
\end{equation}
where $\omega^2 = r^2 + l^2$ and
\begin{equation}
    F(\omega) = 
    \begin{cases}
        \omega^{-\alpha_1},  & \omega < r_f \\
        \frac{1}{c}\omega^{-\alpha_2}, & \omega >= r_f
    \end{cases} .
	\label{eq:F(omega)}
\end{equation}

Here, $I_0$ is a normalisation constant and $\alpha_1$ and $\alpha_2$ are the power law indices. The jump parameter, $c$, quantifies a jump in ICM density at the power law break (that results in a discontinuity in the measured surface brightness profile), with $c = 1$ being the case where there is not a jump. At a shock front, a discontinuity in gas density is expected, so $c > 1$ can indicate the presence of a merger shock. The break in the power law was fixed at the front radius, $r_f$.

\begin{table}
	\centering
	\caption{Best fit parameters for the northern and southern surface brightness profiles (Figure \ref{SB}), using the broken power law model described by Equations \ref{eq:I(r)} and \ref{eq:F(omega)}.}
	\label{tab:SB}
	\begin{tabular}{cccc} % four columns, alignment for each
		\hline
		Profile & $\alpha_1$ & $\alpha_2$ & c\\
        \hline
        North & 0.8 $\pm$ 0.1 & 1.6 $\pm$ 0.3 & 1.1 $\pm$ 0.1\\
        \hline
		  South & 0.47 $\pm$ 0.06 & 1.5 $\pm$ 0.8 & 1.3 $\pm$ 0.1\\
		\hline
	\end{tabular}
\end{table}

The best fit parameters for both profiles are shown in Table \ref{tab:SB}. In both cases, the jump parameter is close to 1, corresponding to a minor jump at the break. In the north, $c$ is consistent with 1 at the 1$\sigma$ level so the jump is not significant, and therefore does not confirm a definitive shock. However, the southern profile has $c \geq 1$ at the 3$\sigma$ level, so its break is significant. This is consistent with the southern edge on the GGM filtered image being a shock front. The Mach number corresponding to the southern density jump is $M = 1.20 \pm 0.09$, when estimated with the Rankine–Hugoniot jump condition,
\begin{equation}
    M = \left[\frac{2c}{\gamma + 1 - c(\gamma - 1)}\right]^\frac{1}{2},
	\label{eq:Mach}
\end{equation}
for a monoatomic gas ($\gamma =$ 5/3). This value likely represents a lower limit on the true Mach number, due to the surface brightness edge potentially being ``smoothed'' by projection effects or point spread function (PSF) blurring. 

%meaning that there is not a significant change in surface brightness at the proposed shock front. This can be seen by the shape of the fits. Typically, one would expect a jump of $c \geq 2$ at a shock front (e.g. \cite{Gutierrez2005}). Because of the lack of a significant jump in both cases, we cannot confirm that the edge features seen in the GGM filtered image represent definitive shocks. 

\subsection{Thermal Structure}

\begin{figure}
\centering
\includegraphics[width=0.45\textwidth]{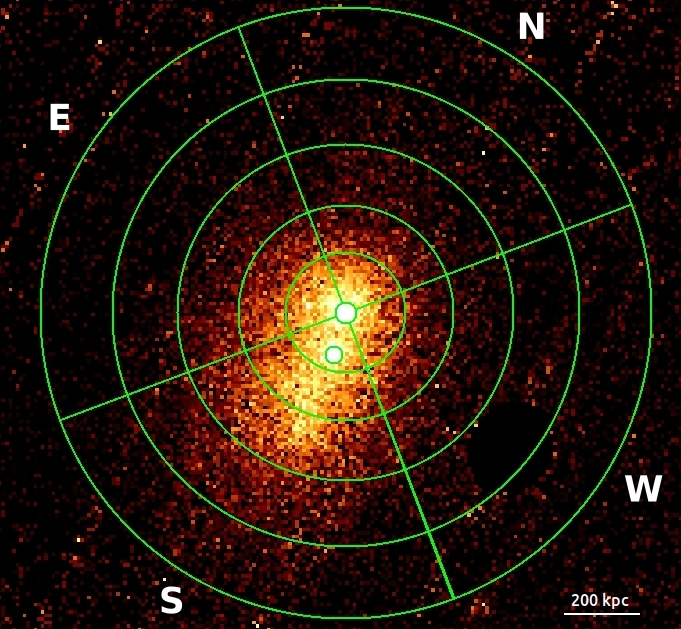}
\caption{\label{Tbins} X-ray image of Abell 272 with the extraction regions used for the temperature profile (Figure \ref{Tprof}) shown in green.}
\end{figure}

\begin{figure}
\centering
\includegraphics[width=0.5\textwidth]{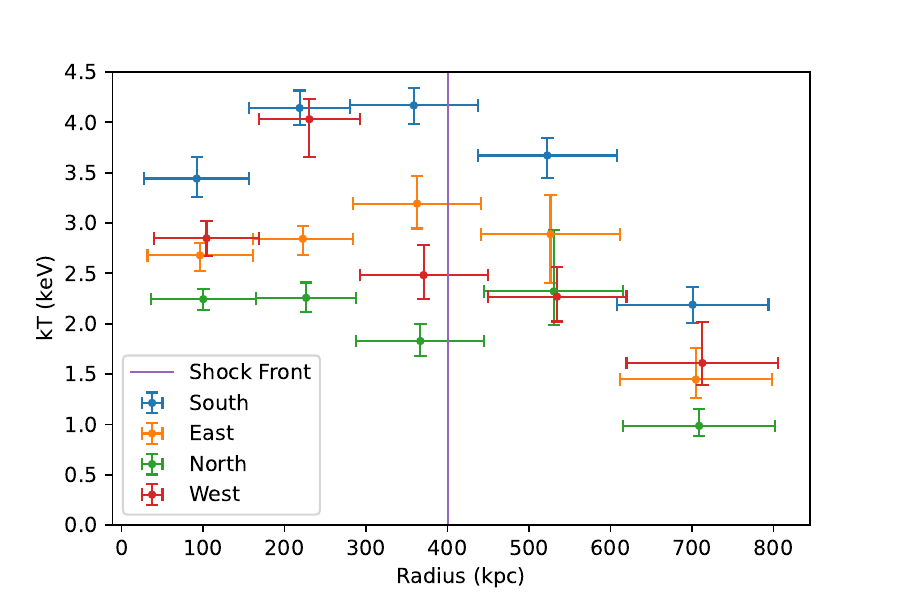}
\caption{\label{Tprof} Temperature profile showing the radial variation in temperature within the four sectors shown in Figure \ref{Tbins}. The four individual profiles have been offset slightly along the x-axis for clarity. The location of the southern shock front is shown by a purple line.}
\end{figure}

\begin{figure*}
    \begin{subfigure}[b]{0.25\textwidth}
    \includegraphics[width=1.2\textwidth]{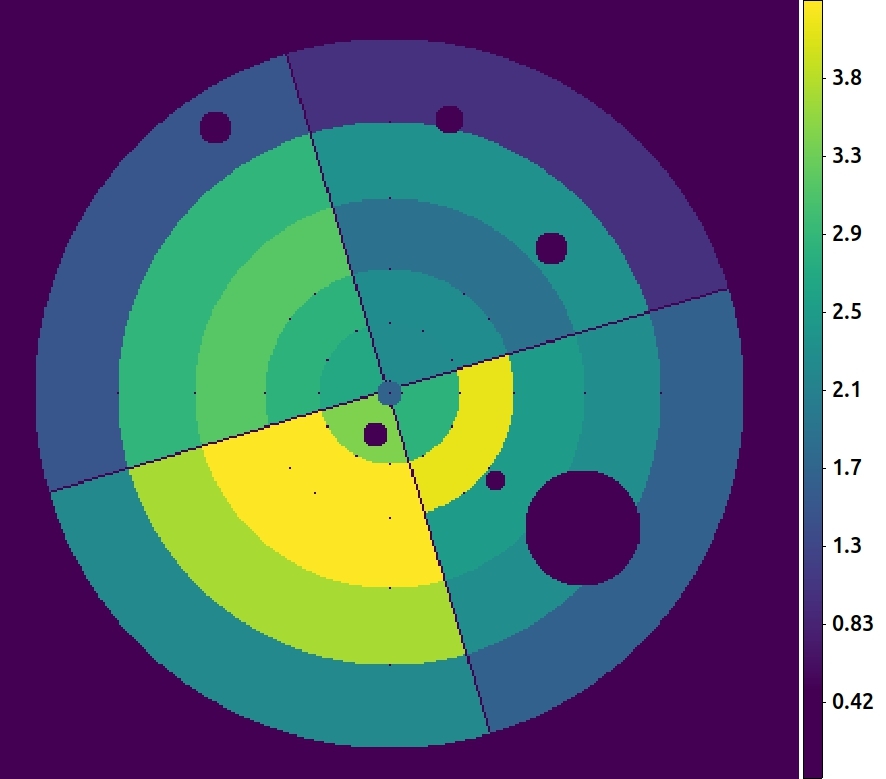}
    \end{subfigure}
    \hspace{8mm}
    \begin{subfigure}[b]{0.25\textwidth}
    \centering
    \includegraphics[width=1.2\textwidth]{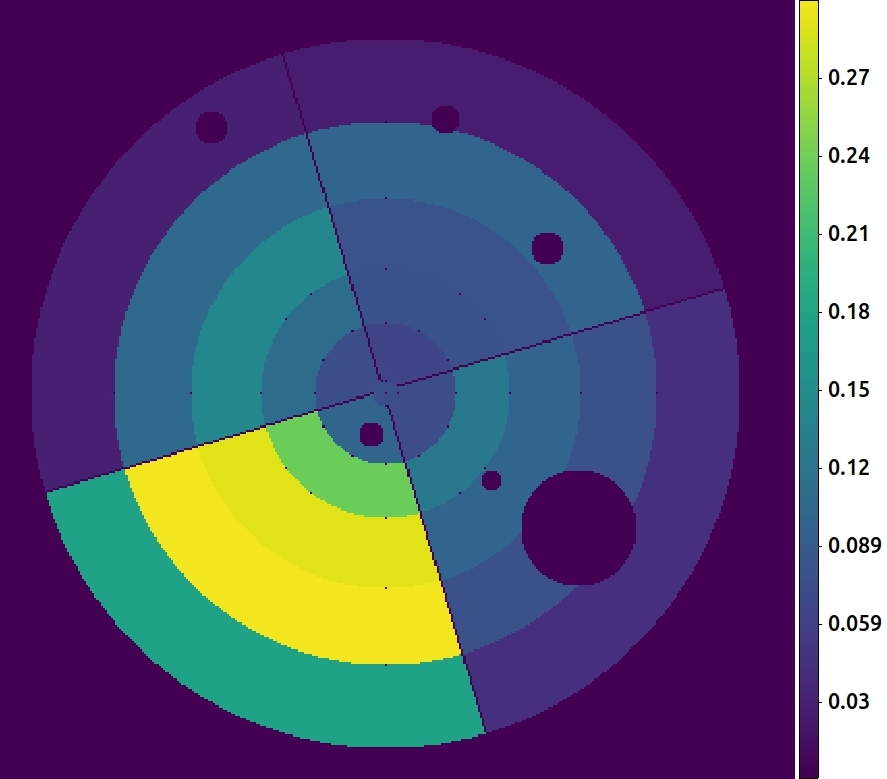}
    \end{subfigure}
    \hspace{8mm}
    \begin{subfigure}[b]{0.25\textwidth}
    \centering
    \includegraphics[width=1.2\textwidth]{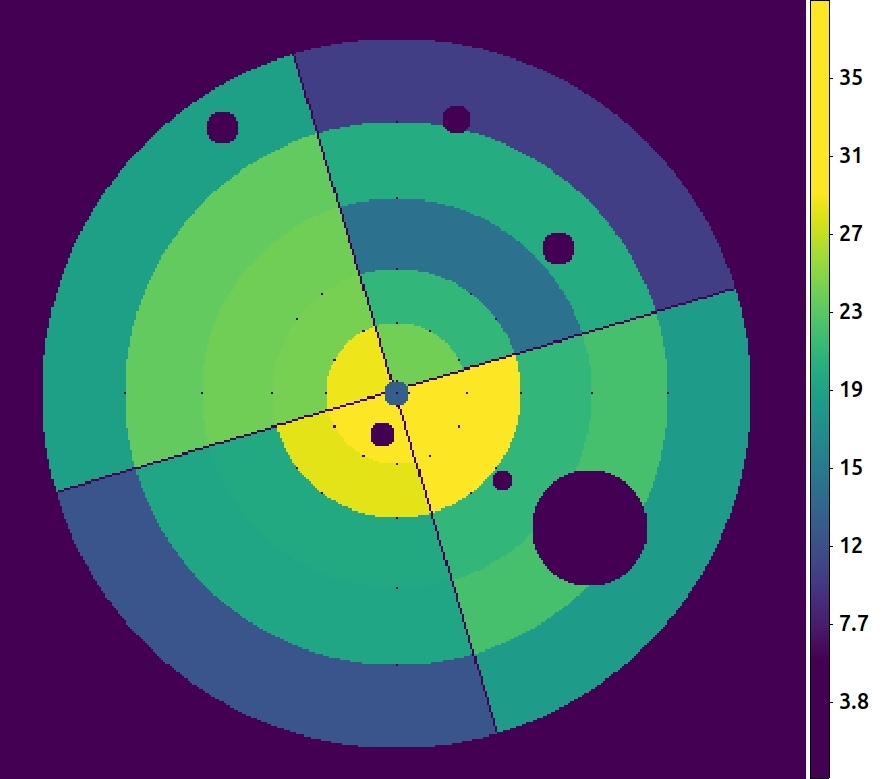}
    \end{subfigure}
\caption{\label{TMap} Temperature (left), pseudo-pressure (centre) and pseudo-entropy (right) maps of Abell 272, all using the bins shown in Figure \ref{Tbins}. Yellow regions represent the highest values, and dark blue the lowest. Circles show removed point sources. The temperature maps is in units of keV, while the pseudo-pressure and pseudo-entropy maps are in arbitrary units.}
\end{figure*}

To examine how the average temperature varied throughout Abell 272, we created temperature profiles in four directions. This was done by splitting the cluster into four sectors, each with five radial bins, and extracting the spectra from each bin. The temperature of each bin was determined by fitting its spectra with an APEC model. %(since kT is a parameter of the model). 
Figure \ref{Tbins} shows the bin regions on the X-ray image. All point sources were excluded from the extracted spectra. Abundances were allowed to vary for each bin, though the uncertainties on the best fit values were too large to draw any conclusions on the variation throughout the cluster. We restricted the fitting range to between 0.1 and 1.0 to prevent best fit abundances that would be unreasonable for emission from the ICM. 

Figure \ref{Tprof} shows the temperature profile along each sector. %, with kT on the y axis and distance from the cluster core on the x axis. 
This is also shown as a temperature map in the left-hand panel of Figure \ref{TMap}. The southern sector, that contains the most elongated emission, is the hottest at all radii when compared to the mean of the three other sectors, at 2$\sigma$ significance. This implies that the ICM gas may have been heated by a shock. The highest temperature bin ($kT = 4.16 \pm 0.19$ keV) lies to the south of the core and the AGN. As this bin contained the putative shock front identified in the surface brightness analysis, we calculated another estimate for the Mach number of the shock, now using the temperature difference between this hot bin and the mean of the three other bins at the same radius (instead of the density jump). Following the Rankine-Hugoniot jump condition for a temperature jump in a monoatomic gas,
\begin{equation}
    \frac{T_2}{T_1} = \frac{5M^4 + 14M^2 - 3}{16M^2},
	\label{eq:MachT}
\end{equation}
this gave $M=1.7 \pm 0.3$. This value is consistent with the Mach number calculated from the density jump at the 1.6$\sigma$ level, so the statistical significance of the difference is low. However, this does suggest that the measured density jump may be lower than its true value, due to projection effects or PSF blurring. We note that temperature measurements can also suffer from the same effects, but their impact is generally less severe.

The temperature map also has a hot bin to the west of the core (the second from the centre) that does not correlate with any features on the X-ray image. We calculated the statistical significance for the difference between this hot bin with the mean of the corresponding bins for the north and east sectors and obtained a value of 3.19$\sigma$. This implies that the difference is significant, but the reason for the higher temperature is unclear. It should be noted that the spectra were not corrected for PSF scattering from bright, nearby regions. Because of this, it is possible that the western region could be biased due to scattering from the bright southern region, causing it to have a higher kT parameter in the spectral fit.

The middle panel of Figure \ref{TMap} shows a pseudo-pressure map in arbitrary units, using the same bins as the temperature map. Pseudo-pressure values for each bin were calculated as $kTA^\frac{1}{2}$ (as in \cite{Randall2016}), where $A$ is the APEC normalisation scaled by the area of the bin. The map shows that the southern sector has a higher pseudo-pressure than the others, again suggesting that there may be shock heated gas in the south. 

The right-hand panel of Figure \ref{TMap} shows a pseudo-entropy map in arbitrary units, using the same bins as the temperature and pseudo-pressure maps. Pseudo-entropy values were calculated as $kTA^{-\frac{1}{3}}$. The map shows that pseudo-entropy is lower in the northern and southern sectors than in the eastern and western sectors. This is in contrast with what one would expect for a merger along the N-S axis; merger turbulence and shocks would be expected to increase entropy along the merger axis \citep[e.g.][]{zuhone2011parameter}. The lower pseudo-entropy could be the result of low entropy gas being stripped from the cluster cores during the merger, forming an extended low entropy region along the merger axis. However, the map alone does not provide enough evidence to confirm this interpretation, and further observations would be needed to determine the significance of the lower pseudo-entropy regions.

\begin{figure}
\centering
\includegraphics[width=0.45\textwidth]{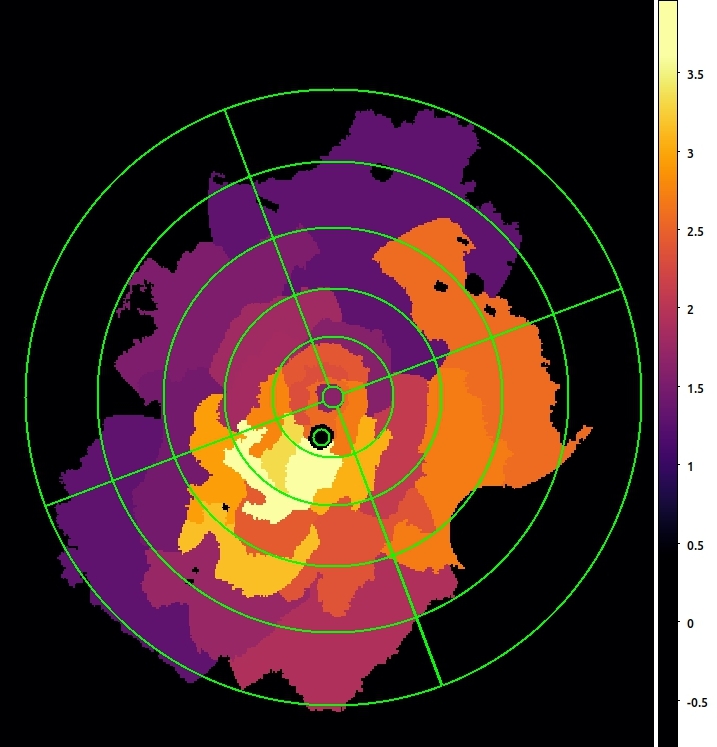}
\caption{\label{Contbin} Contour-binned temperature map of Abell 272. Yellow regions represent the highest temperatures, and purple the lowest. Typical errors on the temperature values are of the order $\pm10\%$. The temperature profile regions from Figure \ref{Tbins} are shown in green.}
\end{figure}

To further investigate the thermal structure of the cluster, we constructed a contour-binned temperature map. This was similar to the previous map (Figure \ref{TMap}, left-hand panel), but the bins were now defined such that each bin had approximately the same X-ray surface brightness throughout. The bins were defined using \verb|contbin| \citep{Sanders2006}. This split the cluster into 29 bins, using a minimum signal to noise of 45 in each bin. %We converted the bins into mask images, in which the value of every pixel inside the bin was 1 and every pixel outside was 0. These images were used to define the regions the X-ray spectra were extracted from in XMM SAS.
As in the case of the temperature profiles, the spectra from each bin were fitted with an absorbed APEC model to determine the average temperature of the bin. 

The resulting temperature map is shown in Figure \ref{Contbin}. As in the temperature profile, the hottest area of the cluster is shown to be south of the core and AGN (the two circular regions in the centre of the map). The highest temperature bin has $kT = 3.97^{+0.51}_{-0.40}$, which is consistent with the hottest bin from the temperature profile in both $kT$ and location. In general, the map and profile bins are consistent with one another. The $1\sigma$ errors on the map bins are of the order $\pm10\%$. One area of apparent disagreement between the profile and map is the hot region to the northwest of the core in the map, which has a higher $kT$ relative to the rest of the cluster in the map than in the profile. However, its temperature is still consistent between the two within their uncertainties. 

\section{Radio Emission}

\begin{figure*}
    \begin{subfigure}[b]{0.25\textwidth}
    \includegraphics[width=1.2\textwidth]{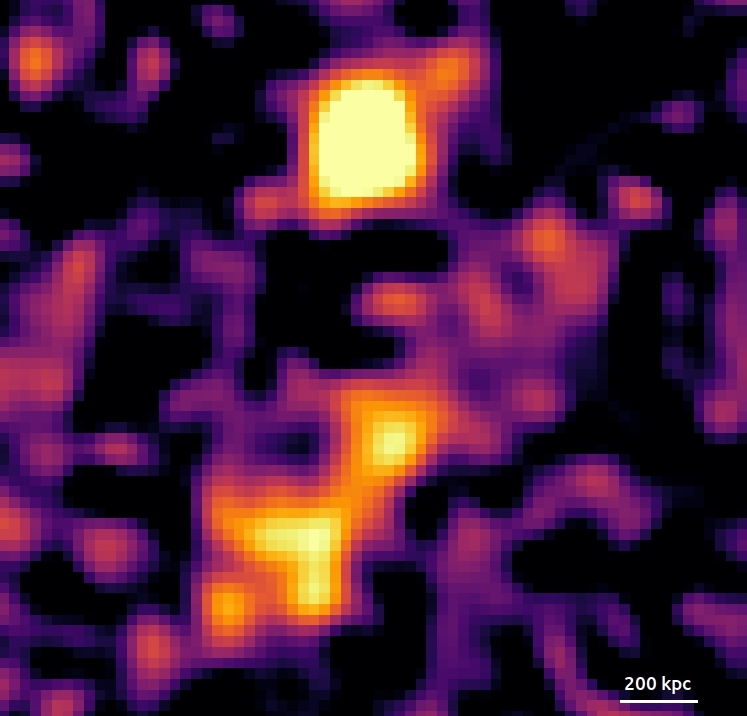}
    \end{subfigure}
    \hspace{8mm}
    \begin{subfigure}[b]{0.25\textwidth}
    \centering
    \includegraphics[width=1.2\textwidth]{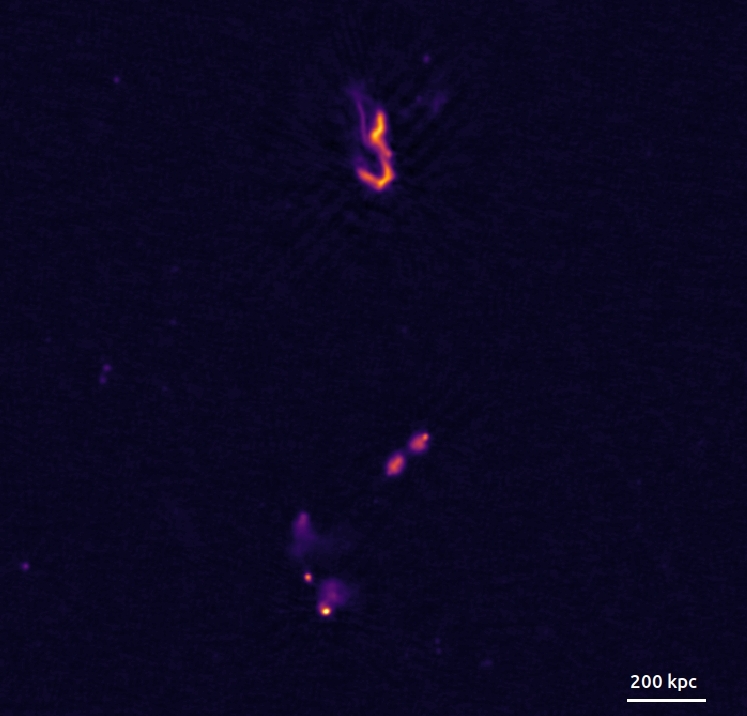}
    \end{subfigure}
    \hspace{8mm}
    \begin{subfigure}[b]{0.25\textwidth}
    \centering
    \includegraphics[width=1.2\textwidth]{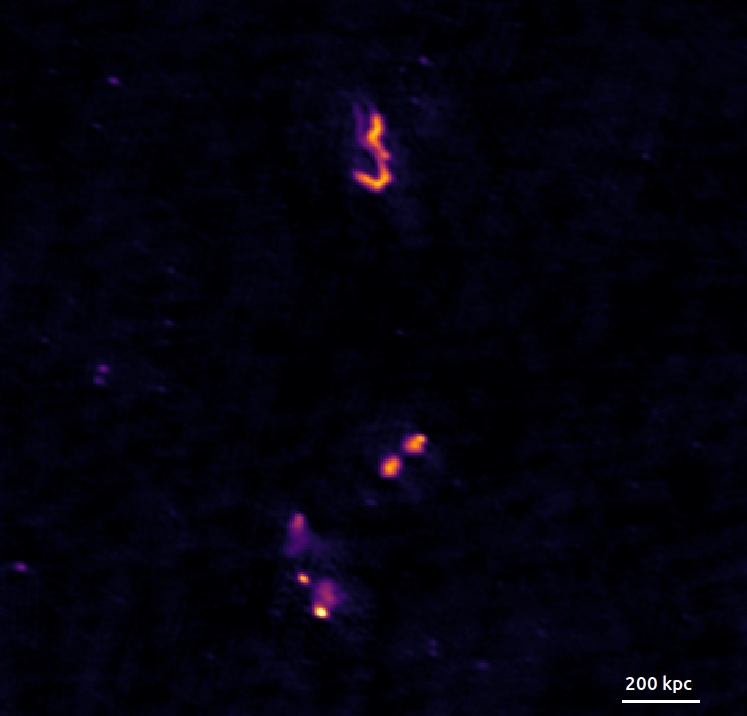}
    \end{subfigure}
\caption{\label{LOFAR} Left: 74 MHz VLSSr image, containing two apparent diffuse radio sources. Centre: 144 MHz LoTSS image, with a diffuse radio source to the north and a set of smaller sources to the south. Right: 325 MHz GMRT image, showing the same sources seen with LoTSS. All three images show the same region of the sky. The angular resolution in the three images is equivalent to 140, 10 and 20 kpc at 74, 144 and 325 MHz, respectively.}
\end{figure*}

\begin{figure}
\centering
\includegraphics[width=0.4\textwidth]{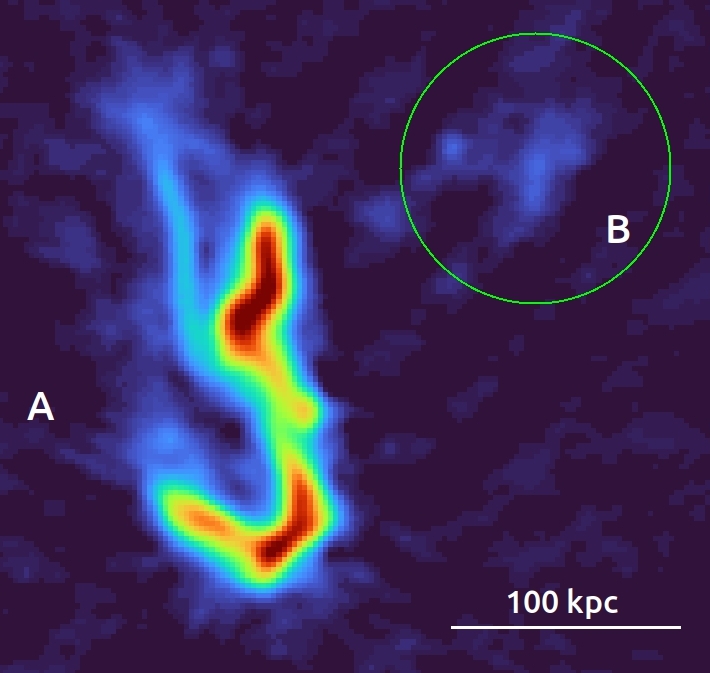}
\caption{\label{USS} Zoomed-in version of the LoTSS image from Figure \ref{LOFAR}, highlighting the northern region. The USS source (A) is shown to the left, with the much fainter AGN detection (B) to its northwest inside the green circle. The brightest regions are shown in dark red, and the fainter regions in blue.}
\end{figure}

\begin{table}
	\centering
	\caption{Summary of radio observations and their associated flux densities.}
	\label{tab:radioflux}
	\begin{tabular}{ccc} % four columns, alignment for each
		\hline
		Observation & Frequency (MHz) & Flux Density (mJy)\\
        \hline
		  VLSSr & 74 & $2310 \pm 450$\\
		\hline
        LoTSS & 144 & $783.1 \pm 78.3$\\
        \hline
        TGSS ADR1 & 147 & $736.4 \pm 74.6$\\
        \hline
        GMRT & 325 & $275 \pm 14$\\
        \hline
        NVSS & 1425 & $7.1 \pm 0.7$\\
        \hline
	\end{tabular}
\end{table}

\begin{figure}
\centering
\includegraphics[width=0.5\textwidth]{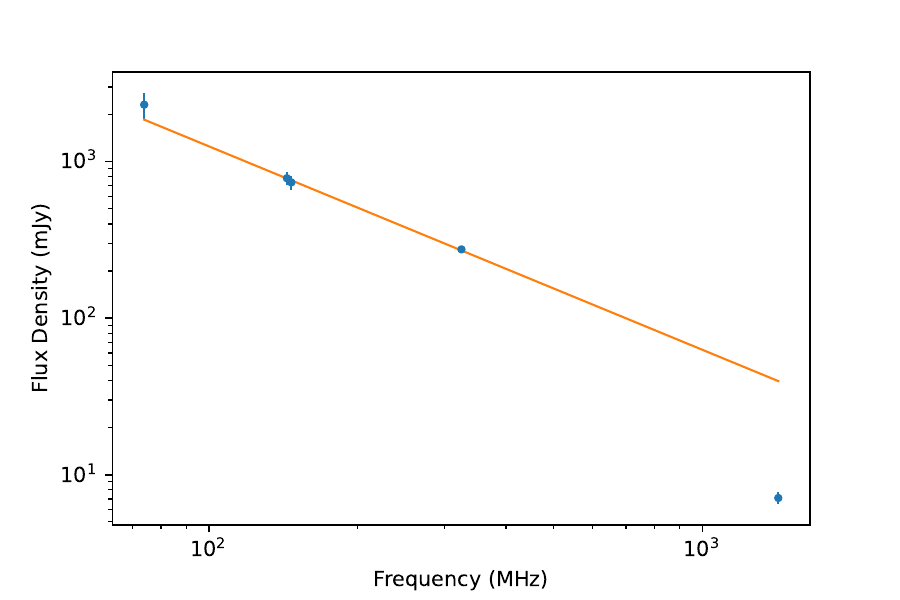}
\caption{\label{radiospec} Radio spectral distribution of the USS source from 74 MHz to 1400 MHz. The orange line shows the best fit radio power law with spectral index $\alpha = -1.3 \pm 0.1$. The spectrum steepens after the 325 MHz measurement.}
\end{figure}

Figure \ref{LOFAR} shows three co-aligned radio images of Abell 272. The left-hand panel is the 74 MHz VLSSr image, the middle panel is the 144 MHz LoTSS image and the right-hand panel is the 325 MHz GMRT image. Two apparent extended radio sources are visible in the VLSSr image - a bright one in the north, and a fainter and more irregular one in the south. 

The high resolution LoTSS and GMRT images show radio emission from an ensemble of small angular size sources to the south. Their morphology indicates that they are double-lobed radio galaxies. One of the radio sources is coincident with a galaxy in the Sloan Digital Sky Survey \citep[SDSS,][]{SDSS} at redshift $z$ = 0.6438 $\pm$ 0.0002, which is significantly different from the cluster redshift ($z$ = 0.0877). We therefore make the assumption that the collection of galaxies is unrelated to Abell 272.

The northern emission is shown to be a pair of sources in the LoTSS and GMRT images, which we label A and B in Figure \ref{USS}. Source A is diffuse and very bright. Source B is smaller and much fainter, located to the northwest of A. B is located at the same position as the AGN detected in the X-ray image as the southern bright peak (coincident with G6), and is therefore most likely its radio counterpart. Its separation from source A suggests it is unrelated. 

The LoTSS and GMRT images provide a much clearer look at the structure of source A; it has an irregular morphology that is consistent with double-lobed radio jets, akin to ones associated with AGN. The two bright lobes have an S-shaped morphology. Fainter filaments can be seen that split off from the main lobes, giving the source a complicated structure. %The measured fluxes of the source are given in Table ???. 

Source A was originally classified as a USS source due to its measured spectral index of $-1.9 \pm 0.1$ between the frequencies 74 MHz and 1.4 GHz (calculated using the archival VLSSr and NVSS flux densities). Its full spectral distribution from 74 MHz to 1.4 GHz with the additional measured flux densities (which are listed in Table \ref{tab:radioflux}) is shown in Figure \ref{radiospec}. The data is described well by a power law of the form $F_\nu \propto \nu^\alpha$ with $\alpha = -1.3 \pm 0.1$ from 74 MHz to 325 MHz. Beyond this range, the spectrum steepens, as illustrated by the much lower flux at 1.4 GHz.

\begin{figure}
\centering
\includegraphics[width=0.45\textwidth]{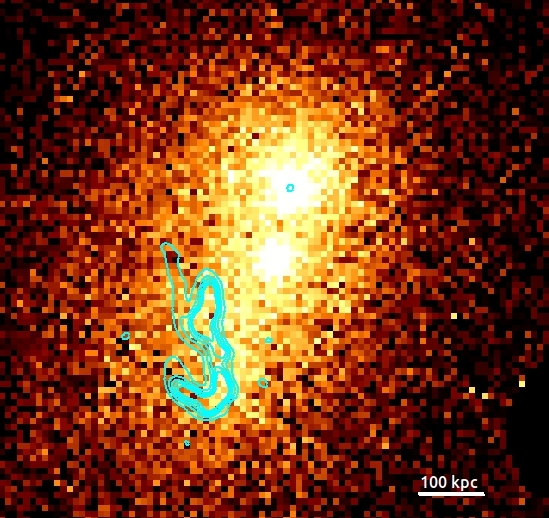}
\caption{\label{RadioCon} X-ray image of Abell 272 with LoTSS contours (from Figure \ref{LOFAR}) overlaid in blue, showing its location in the southern extended region of the ICM.}
\end{figure} 

The USS source is located in the extended part of the ICM south of the core. Its exact position is shown in Figure \ref{RadioCon}, in which the contours from the LoTSS image (Figure \ref{LOFAR}) are overlaid on the X-ray image of the cluster (Figure \ref{X-ray}).

\begin{figure*}
\centering
\begin{subfigure}[b]{0.4\textwidth}
    \centering
    \includegraphics[width=1.1\textwidth]{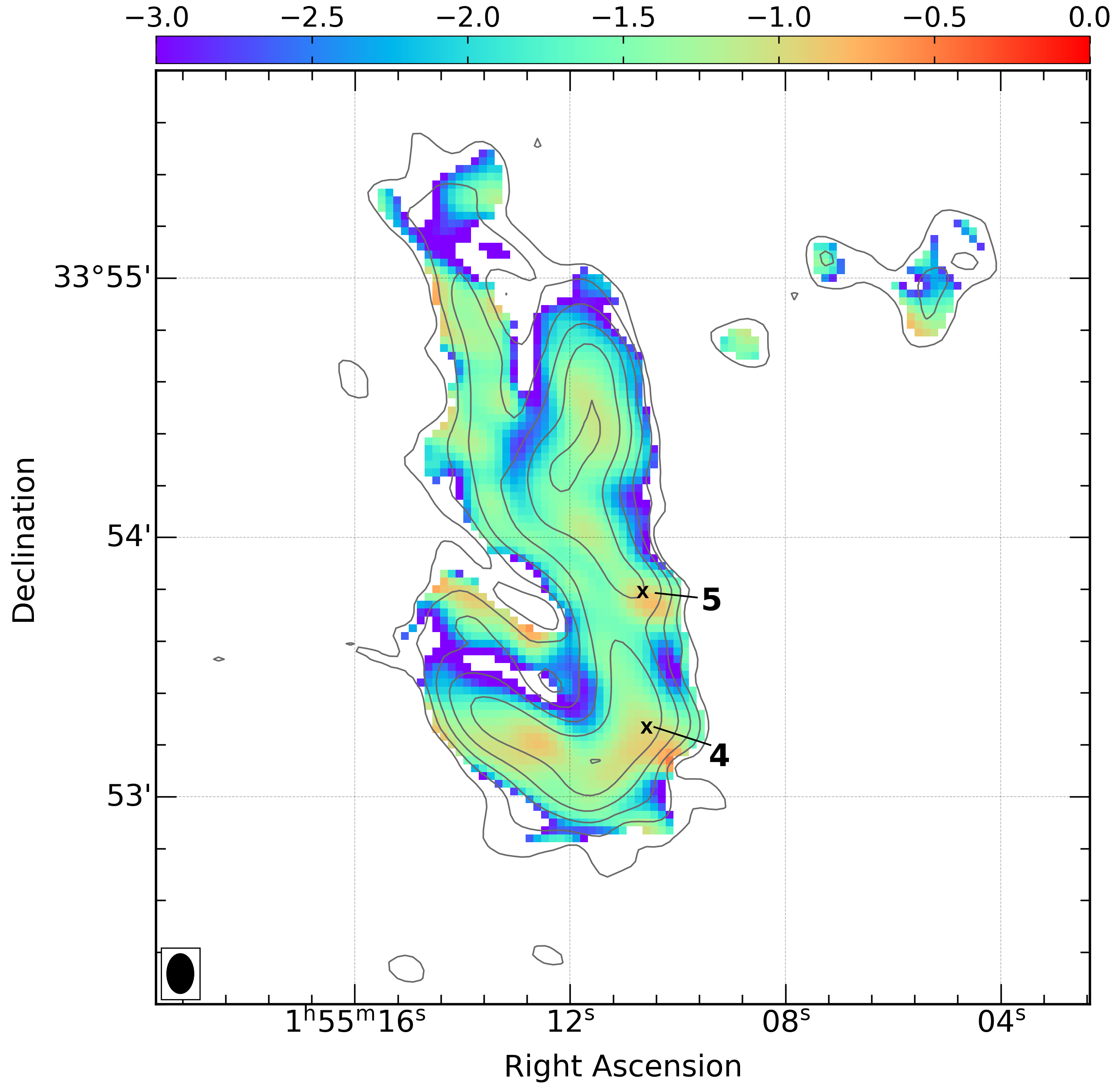}
    \end{subfigure}
    \hspace{8mm}
    \begin{subfigure}[b]{0.4\textwidth}
    \centering
    \includegraphics[width=1.1\textwidth]{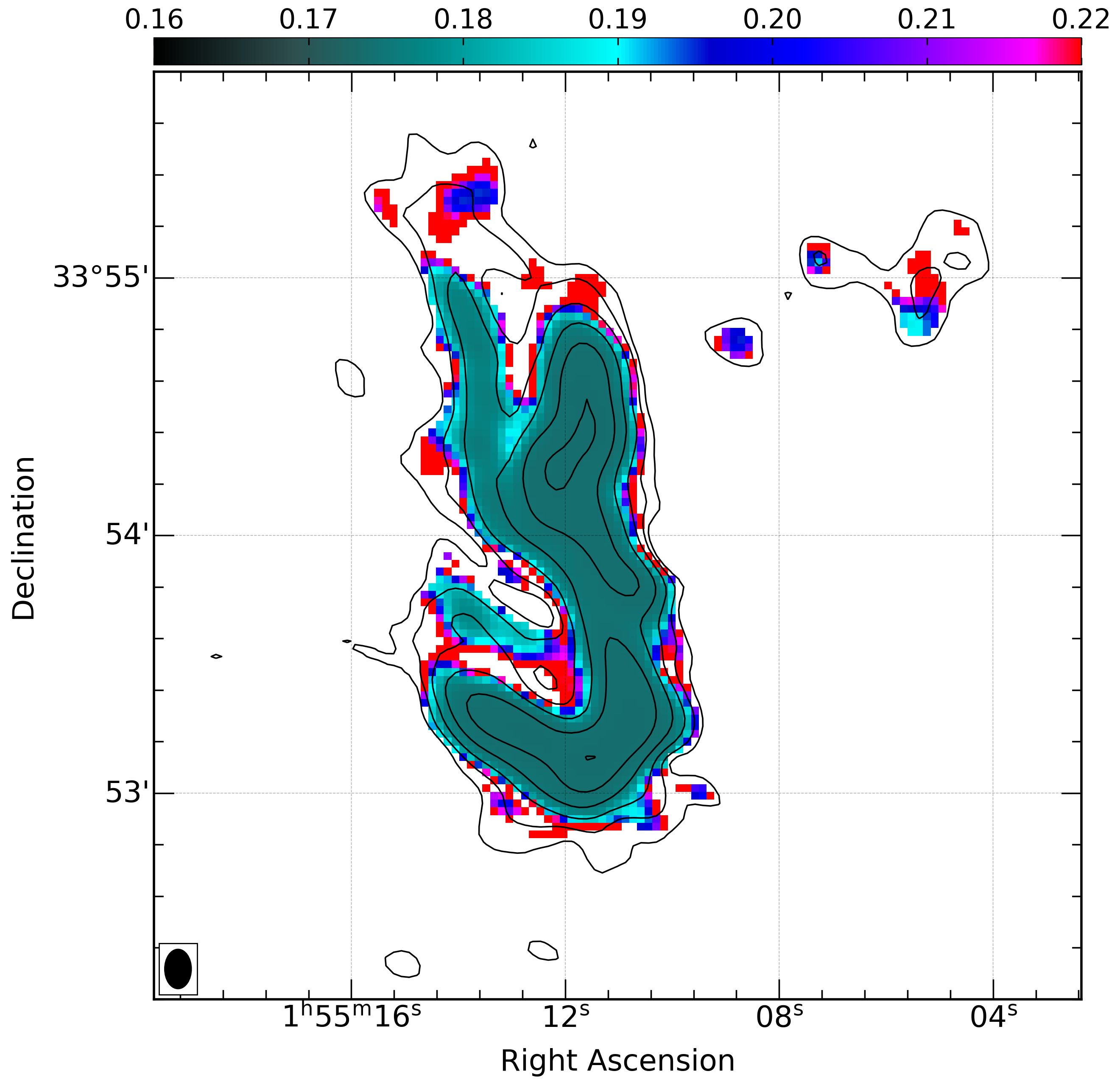}
    \end{subfigure}
\caption{\label{SpecIndex} Left: Spectral index map of the USS source between 144 MHz and 325 MHz. The 4 and 5 indicate the galaxies from Figure \ref{OpticalIm} that coincide with maximum points on the map. Right: Error map for the same frequency range. LoTSS radio contours are overlaid on both maps.}
\end{figure*} 

A spectral index map of the source is shown in the left-hand panel of Figure \ref{SpecIndex} for the frequency range 144 MHz $\leq \nu \leq$ 325 MHz. The right-hand panel shows the uncertainties across the map, which are consistently around $\pm 0.17$ for the majority of the radio structure. The map contains three emission peaks, of which two coincide with galaxies (4 and 5 from Figure \ref{OpticalIm}). Radio-emitting galaxies would typically be expected to fall within regions of flatter spectral index, so these two galaxies could be linked to the radio emission, although we note that their locations do not perfectly match those of the spectral index maximums.

\section{Discussion}

\subsection{Cluster Dynamical State}

\begin{table}
	\centering
	\caption{Spectroscopic and photometric redshifts for the 16 galaxies labelled in Figure \ref{OpticalIm}. Dashes indicate any values that are unavailable.}
	\label{tab:redshift}
	\begin{tabular}{ccc} % four columns, alignment for each
		\hline
		Galaxy & Spectroscopic $z^a$ & Photometric $z^b$\\
        \hline
		  1 & 0.0869 & $0.114 \pm 0.0241$\\
		\hline
        2 & 0.0893 & $0.107 \pm 0.0254$\\
        \hline
        3 & 0.0887 & $0.138 \pm 0.0138$\\
        \hline
        4 & 0.0882 & $0.102 \pm 0.0136$\\
        \hline
        5 & 0.0877 & $0.108 \pm 0.0137$\\
        \hline
        6 & 0.0880 & $0.120 \pm 0.0269$\\
        \hline
        7 & - & $0.118 \pm 0.0128$\\
        \hline
        8 & 0.0893 & $0.087 \pm 0.0090$\\
        \hline
        9 & 0.0856 & $0.101 \pm 0.0117$\\
        \hline
        10 & 0.0894 & $0.109 \pm 0.0196$\\
        \hline
        11 & 0.0849 & -\\
        \hline
        12 & 0.0901 & $0.091 \pm 0.0124$\\
        \hline
        13 & 0.0926 & $0.068 \pm 0.0220$\\
        \hline
        14 & 0.0831 & $0.091 \pm 0.0084$\\
        \hline
        15 & 0.0838 & $0.114 \pm 0.0195$\\
        \hline
        16 & - & $0.132 \pm 0.0512$\\
        \hline
	\end{tabular}
    \begin{tablenotes}
        \item $^a$ Cited from \cite{sargent1972cluster}.
        \item $^b$ Cited from \cite{SDSS}.
    \end{tablenotes}
\end{table}

\begin{figure}
\centering
\includegraphics[width=0.45\textwidth]{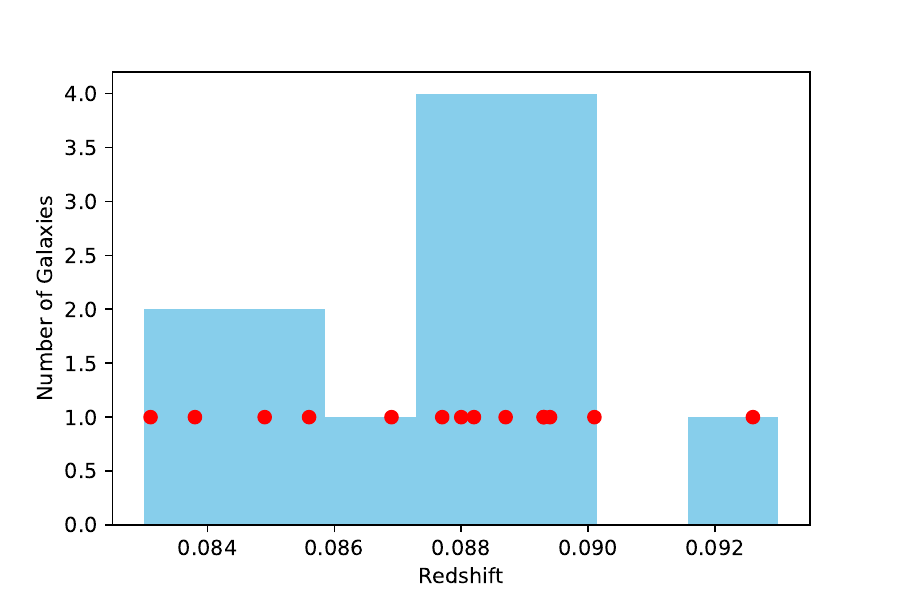}
\caption{\label{Optical} Histogram of spectroscopic redshifts for 14 cluster members, binned bewteen $z = 0.083$ and $z = 0.093$ with bin widths of 0.0014. The individual values for each galaxy are shown as red points.}
\end{figure} 

The X-ray observations of Abell 272 (Figure \ref{X-ray}) suggest that it is a cluster merger, with the merger axis in the N-S direction. This is evident in the ICM emission, which is extended along the merger axis to the south of the core. The extended region has a higher temperature than the rest of the cluster (at 2$\sigma$ significance), suggesting that the gas may have been heated by a merger shock. 

Two potential shock fronts are visible in the GGM filtered image (Figure \ref{GGM}). One is located north of the core, and the other to its south. Both fronts appear to have propagated along the proposed merger axis. We find that only the surface brightness discontinuity in the south is significant (at the 3$\sigma$ level), and conclude that it most likely represents a shock front. The modelled density jump is relatively small ($c = 1.3 \pm 0.1$) and corresponds to a lower Mach number than the temperature jump at 1.6$\sigma$ significance ($M_\rho = 1.20 \pm 0.09$ and $M_T = 1.7 \pm 0.3$), so
%However, we find no significant jump in surface brightness across either shock front, which one would typically expect to see. Therefore, it is uncertain whether these are shock fronts or not.
it is possible that the direction of the merger may have a line of sight component in addition to the N-S motion implied by the X-ray image. If this is the case, any merger-induced shocks would propagate partially along the line of sight, which could explain why we do not see major jumps in X-ray surface brightness. %This is probable given the image's relatively short and shallow exposure. 

To investigate a possible line of sight component in the merger axis, we analysed the redshifts of various cluster member galaxies; we used the 14 spectroscopic redshifts from \cite{sargent1972cluster}, as well as a set of 35 SDSS photometric redshifts \citep{SDSS}.

The 14 spectroscopic redshifts are shown in a histogram in Figure \ref{Optical}, with the individual values listed in Tabel \ref{tab:redshift}. The majority of the values are between 0.087 and 0.090, including the BCG (with $z = 0.089$). The galaxies that do not fall within this region are those numbered 9, 11, 13, 14 and 15 in Figure \ref{OpticalIm}. Given that these galaxies are spread throughout the cluster, the data seem to imply that there is no simple link between galaxy redshift and location. 

Using the photometric SDSS values, we found that the vast majority of galaxies within the cluster’s virial radius had higher redshifts than the spectroscopic values, approximately between $z$ = 0.10 and $z$ = 0.15. However, the uncertainties on the photometric values are high (most are more than $\pm$0.01), and the values for some of the galaxies that already have spectroscopic redshifts are not consistent, despite the large uncertainties (see Table \ref{tab:redshift}). We therefore choose to focus on the spectroscopic values, since spectroscopic measurements are typically more accurate than photometric ones. This decision reduces the redshift range to $0.083 \leq z \geq 0.093$, corresponding to a maximum relative velocity between galaxies (along the line of sight) of approximately 3000 kms$^{-1}$, which is consistent with the merger velocity of other major mergers \citep[e.g][]{milosavljevic2007cluster}. This velocity difference could also be explained by the presence of multiple structures seen in projection. However, we are unable to make any firm conclusions on either possibility due to the small number of spectroscopic redshifts, and their significant disagreement with the photometric values. 

%Because the range of redshift values is so large for galaxies that appear to be associated with the cluster, there could be multiple structures projected along the line of sight, potentially as part of a large scale cosmic filament. However, we cannot draw any firm conclusions due to the significant disagreement between the spectroscopic and photometric values for several cluster members.

More evidence that multiple structures exist within the cluster arises when our proposed BCG is reexamined; following our initial analysis of the X-ray image (Figure \ref{X-ray}), we chose G8 as the central BCG and identified G6 as a potential second BCG. However, the optical image (Figure \ref{OpticalIm}) shows that G6 is smaller and has a fainter magnitude than one would typically expect for a BCG when compared to other cluster members (SDSS r band magnitude = 17.08), whereas G5 exhibits the expected characteristics (r = 15.57), as does G8 (r = 15.68). These two galaxies are brighter than all other cluster members by at least 0.6 in the r band. Therefore, it is likely that galaxies 5 and 8 are BCGs, and G6 is merely a cluster member hosting a bright X-ray AGN. In this interpretation, G5 and G8 are the BCGs of two merging subclusters (which we will henceforth refer to as the southern and northern subclusters), each with their own ICM component. The difference in spectroscopic redshift between the two BCGs is only $\Delta z=0.0016$ (corresponding to a velocity difference of 456 km s$^{-1}$), so it is unlikely that the two structures are seen in projection. 

Because the northern subcluster has a cool core, it has not yet been heavily disrupted by the merger. In contrast, the southern subcluster does have a non-cool core and therefore has already been disrupted by a merger event. If this disruption was only due to the merger between the two subclusters we would typically expect to find that both have non-cool cores, so the southern subcluster has likely been disturbed by a different merger event. Given that non-cool cores have lifetimes comparable to the age of the universe (as a result of their low central gas density), this other merger could have happened a long time ago. Alternatively, it could be a recent (or ongoing) merger along the line of sight, and hence cannot be easily observed. There is also a possibility that the disruption of the southern subcluster is solely due to the interaction with the northern subcluster; while this is less common, there are examples of small subclusters retaining their cool core in a merger with a larger structure, such as in the Bullet Cluster \citep{2006ESASP.604..723M}. Therefore, this may be a similar merger in which the northern subcluster is a small, low mass "bullet" that has passed through the larger, high mass southern subcluster, leaving the latter in a disturbed state.

Interpreting A272 as a system with two merging subclusters also provides another explanation for the high temperature region to the south of the cool core; since this region contains the BCG of the southern subcluster (G5), the high temperature could simply be due to the non-cool core. This temperature would require the southern subcluster to be the more massive of the two, as in the "bullet" merger scenario discussed above. Higher angular resolution X-ray observations and more spectroscopic redshift measurements would be needed to determine if this is actually the cause of the high temperature, instead of the shock-heating of the ICM.

%Using the photometric SDSS values, we found that the vast majority of galaxies within the cluster’s virial radius had higher redshifts than the BCG, approximately between $z$ = 0.10 and $z$ = 0.15. This difference is too great to be due to the peculiar velocity of the BCG; the difference between $z$ = 0.15 and $z$ = 0.087 gives $v \approx$ 18900 km s$^{-1}$. Because of this, we made the assumption that these galaxies are associated with the cluster, despite the discrepancy with the BCG redshift. We note that the Two Micron All Sky Survey \citep[2MASS,][]{2MASS} gives the BCG redshift as $z$ = 0.10 $\pm$ 0.01, which is more consistent with the surrounding galaxies. Figure \ref{Optical} shows plots of redshift against position for the 35 galaxies. The left-hand panel displays redshift against right ascension and the right-hand panel shows redshift against declination. The BCG is marked as an orange point (using the SDSS value). It is clear from the plots that the errors on these redshifts are too large for any trends to be identified. However, the fact that the range of redshift values is so large for galaxies that appear to be associated with Abell 272 means that there could be an undetectable structure along the line of sight, potentially as part of a large scale structure filament.

\subsection{The Nature of the USS Radio Source}

\begin{figure}
\centering
\includegraphics[width=0.45\textwidth]{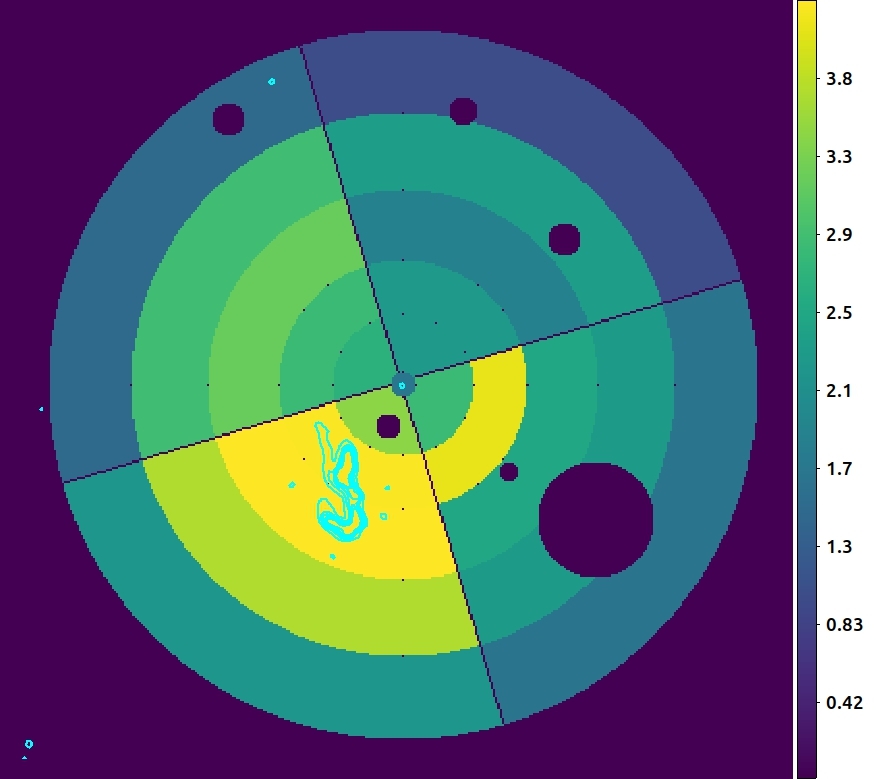}
\caption{\label{TMapCon} Temperature map of Abell 272 (Figure \ref{TMap}) with LoTSS contours (from Figure \ref{LOFAR}) overlaid in blue.}
\end{figure}

The USS source is located in the bright, extended part of the ICM south of the core (as shown in Figure \ref{RadioCon}). This is also the hottest region of the cluster; Figure \ref{TMapCon} shows the temperature map (from Figure \ref{TMap}) with the LoTSS radio contours overlaid in blue. It can be seen that the USS source falls within the two highest temperature bins. This is consistent with the radio source being re-energised by a shock due to the cluster merger. 

It is unlikely that the USS source is a radio relic because the double-lobed morphology contrasts with typical relics, which are usually extended along a shock front. %Furthermore, the source is roughly 200 kpc long, whereas typical relics are greater than 1 Mpc. 
The complex morphology also rules out the possibility that the source is a radio halo, despite being centred on a potential BCG (G5), because halos typically have a far more uniform structure. 

The double-lobed structure of the USS source implies that it is likely an AGN relic or radio phoenix. The two lobes appear to be connected to one another in the radio image (Figure \ref{LOFAR}), implying that they may have originated from an AGN at their centre. In the X-ray image, there is a small group of bright pixels at this location that could potentially be an AGN. To determine the statistical significance of this feature, we compared its surface brightness with the average surface brightness in that region of the ICM. We found the potential source ($r=7$ kpc) to have a surface brightness of $1.52 \pm 0.25$ cts/arcsec$^2$, while the mean surface brightness in an annulus around it (7 kpc $< r <$ 38 kpc) was $0.78 \pm 0.02$ cts/arcsec$^2$. This meant that the difference was significant at the 3$\sigma$ level. Furthermore, the proposed BCG of the southern subcluster lies at the centre of the S-shaped source (G5 in Figure \ref{OpticalIm}), so it is likely that the radio lobes originated from the central AGN of the subcluster. The region of the ICM in which the source is located seems to have likely been heated by a merger shock, that could have caused adiabatic compression of the ICM and re-energised the AGN lobes. We also note that G5 is likely the BCG of a non-cool core cluster, which rarely have radio lobes as bright as those observed, due to the lack of cool gas to feed the AGN. This gives additional evidence for the re-energisation of the lobes, as they would likely be fainter otherwise. The structure of the northern lobe, which appears to be split, also supports this; one half could be the radio lobe directly from the AGN, while the other is the remnant of a previous radio outburst that was re-energised in a similar manner to a radio phoenix. 

An alternate possibility for the formation of the radio structure can be inferred from the spectral index map of the source (Figure \ref{SpecIndex}). Two of the maximum points on the map are coincident with cluster member galaxies (G4 and G5). While the maximum at G5 further reinforces the interpretation discussed above, the maximum at G4 implies that there could be a second AGN associated with the observed radio emission. If both galaxies hosted AGN, then the USS source could be a mixture of their radio outbursts. This could also explain the splitting of the source, with the filaments being overlapping tails of the two radio galaxies seen in projection, thus giving the impression of an overall S-shaped morphology. We note that the cluster merger likely still impacted the radio emission in this case, due to the location of the source within the cluster. 

\subsection{Inverse Compton Emission}

In theory, there should be some inverse Compton (IC) emission at the location of the USS source, due to the interaction between the synchrotron radio emitting electrons from the source and the cosmic microwave background (CMB). IC emission in clusters is typically difficult to isolate from the prevalent thermal bremsstrahlung emission \citep[e.g.][]{Wik2014}, although a $4.6\sigma$ detection has recently been reported in a galaxy group associated with the radio source MRC 0116+111 \citep{mernier2023discovery}.

%Despite this prediction, the presence of IC emission in galaxy clusters has yet to be conclusively detected (see \cite{Wik2014} for recent findings). 

To search for possible IC emission, we analysed the X-ray spectra extracted from an ellipse encompassing the USS source. We replaced the APEC component of our spectral model with an absorbed power law component to model the IC emission from the source \citep{Sarazin1999}. In doing this, we made the assumption that all X-ray emission in the region is due to IC emission. This is clearly incorrect, as there would still be some thermal bremsstrahlung emission, but by making this assumption we were able to find the maximum possible IC flux. We tested an additional, more realistic model with both bremsstrahlung and IC components, but their parameters could not be constrained. Therefore, we focus solely on the model without the bremsstrahlung component. We note that the background components were not changed from the original model outlined in Section 2.2.

Statistically, the IC model was a negligible improvement to the fit when compared to the thermal model outlined in Section 2.2. The F-test probability was 29\%, which is much higher than the probability of 5\% or less that indicates a significant improvement. The best fit spectral index $\Gamma$ was $-1.3^{+0.2}_{-0.7}$, meaning that the IC component was not well constrained in the fit. This value does match the best fit radio spectral index ($\alpha^{\text{74 MHz}}_{\text{325 MHz}} = -1.3$), which is what would typically be expected for emission from the same electron population \citep{Pacholczyk1977}. The model gave an IC flux of $2.2^{+0.2}_{-0.2}\times10^{-13}$ ergs cm$^{-2}$ s$^{-1}$.

The IC flux can be used to determine a lower limit on the ICM magnetic field strength \citep{Petrosian2001}:
\begin{equation}
    B = \left(\frac{20 \text{keV}}{kT}\right) \left(\frac{\nu}{\text{GHz}}\right)^{(p-1)/(p+1)}e^{\frac{2.84(p-r)}{p+1}} \mu G
	\label{eq:B}
\end{equation}
where r is given by
\begin{equation}
    r = 0.7\text{ln}\left[\frac{R_{obs}(kT,\nu)}{1.11  \times10^{-8}}\right].
	\label{eq:r}
\end{equation}

$p$ is the power law slope of the electron energy distribution ($N(E) \propto E^{-p}$) and is related to $\Gamma$ by $\Gamma = (p+1)/2$. $R_{obs}(kT,\nu)$ is the ratio between the IC flux at temperature $kT$ and the radio flux at frequency $\nu$. For $\Gamma$ = $1.3$, this gave a magnetic field strength of 0.03 $\mu$G. %This is almost 100 times smaller than the $B = 1.9\pm0.3$ $\mu$G detection by \cite{mernier2023discovery}.

Because the addition of an IC component in the spectral model does not give a significant improvement in the fit, and IC emission is difficult to detect in galaxy clusters, we conclude that it is more likely that the X-ray emission in this region is purely thermal bremsstrahlung emission. The calculated magnetic field strength is, therefore, a conservative lower limit and the IC flux an upper limit, since it is unlikely that much IC emission is present. 

\section{Conclusion}

In this paper, we presented X-ray and radio observations of the galaxy cluster Abell 272. The X-ray image shows a disturbed system, elongated from north to south, that is consistent with Abell 272 being an ongoing major cluster merger. The X-ray temperature profile reveals a hot region south of the cluster’s cool core, suggesting the presence of shock-heated gas in the ICM. Analysis of the surface brightness provides evidence of a merger shock front in the south of the system at the $3\sigma$ level, with a corresponding Mach number $M = 1.20 \pm 0.09$. The Mach number was also calculated from the temperature jump, giving $M = 1.7 \pm 0.3$. We investigated the possibility of the merger axis having a component along the line of sight, but there is currently not enough precise redshift data available to verify this suggestion. After comparing optical and X-ray images, galaxies 5 and 8 (from Figure \ref{OpticalIm}) were identified as the potential BCGs of two merging subclusters. Due to the northern and southern subclusters having a cool core and non-cool core respectively, we proposed two merger scenarios: either the southern subcluster has been disturbed by some other merger event, and is now merging with the northern subcluster; or the northern subcluster is much less massive than the southern subcluster, and the former has retained a cool core after piercing through the latter in a similar manner to the Bullet Cluster. 

Radio observations show a USS diffuse radio source in this hot region, with spectral index $\alpha^{\text{74 MHz}}_{\text{325 MHz}} = -1.3 \pm 0.1$ and steepening at higher frequencies. Given its location, it is likely that the radio emission is a result of normal-thermal particles that have been re-energised by a shock. The source exhibits a double-lobed morphology with a complex filamentary structure, with properties of both AGN relics and radio phoenices. We give two possible interpretations of the source. The first is that the source originated from an AGN (in galaxy 5) at the centre of the two most prominent lobes, and was re-energised via adiabatic compression to produce the observed filamentary morphology. The second interpretation is that there is also a second AGN in galaxy 4 towards the south of the radio source (motivated by a maximum in spectral index at the location of the galaxy), and the complex structure of the source is primarily due to the overlapping radio outbursts from both AGN. Given the current radio data, it is unclear which of the two explanations is most likely, and deeper observations that better resolve the complex structure would be required in order to distinguish between them. 

%Because of the complex filamentary structure, we propose that the USS source could be a radio phoenix caused by the adiabatic compression of an electron population left over from a previous AGN outburst, with the fainter filaments being from older outbursts than the brighter ones. 

Finally, the X-ray observations were used to investigate the presence of IC emission at the location of the USS source. We calculated a conservative upper limit on the theoretical IC flux of $2.2\times10^{-13}$ ergs cm$^{-2}$ s$^{-1}$ and lower limit on the ICM magnetic field strength of 0.03 $\mu$G.

\section*{Acknowledgements}

We are thankful to Tony Mroczkowski, Ewan O’Sullivan and Gerrit Schellenberger for useful discussions.
Support for this work was partially provided by the XMM-Newton grant 80NSSC22K0568. 
AW acknowledges support from STFC grant ST/Y509486/1.
Support for SWR was provided by the Chandra X-ray Center through NASA contract NAS8-03060, and the Smithsonian Institution.
Basic research in Radio Astronomy at the U.S. Naval Research Laboratory is supported by 6.1 Base funding. 
RJvW acknowledges support from the ERC Starting Grant ClusterWeb 804208.
WF acknowleges support from the Smithsonian Institution, the Chandra High Resolution Camera Project through NASA contract NAS8-0306, NASA Grant 80NSSC19K0116, and Chandra Grant GO1-22132X.
ACE acknowledges support from STFC grant ST/P00541/1.
ELB was partially supported by the National Aeronautics and Space Administration, through Chandra Award Number GO2-23123X.
LL acknowledges financial contribution from the INAF grant 1.05.12.04.01. 

\section*{Data Availability}

Supporting research data are available on reasonable request from the corresponding author A. Whyley.

\bibliographystyle{mnras}
\bibliography{sample}

\begin{thebibliography}{}
\makeatletter
\relax
\def\mn@urlcharsother{\let\do\@makeother \do\$\do\&\do\#\do\^\do\_\do\%\do\~}
\def\mn@doi{\begingroup\mn@urlcharsother \@ifnextchar [ {\mn@doi@}
  {\mn@doi@[]}}
\def\mn@doi@[#1]#2{\def\@tempa{#1}\ifx\@tempa\@empty \href
  {http://dx.doi.org/#2} {doi:#2}\else \href {http://dx.doi.org/#2} {#1}\fi
  \endgroup}
\def\mn@eprint#1#2{\mn@eprint@#1:#2::\@nil}
\def\mn@eprint@arXiv#1{\href {http://arxiv.org/abs/#1} {{\tt arXiv:#1}}}
\def\mn@eprint@dblp#1{\href {http://dblp.uni-trier.de/rec/bibtex/#1.xml}
  {dblp:#1}}
\def\mn@eprint@#1:#2:#3:#4\@nil{\def\@tempa {#1}\def\@tempb {#2}\def\@tempc
  {#3}\ifx \@tempc \@empty \let \@tempc \@tempb \let \@tempb \@tempa \fi \ifx
  \@tempb \@empty \def\@tempb {arXiv}\fi \@ifundefined
  {mn@eprint@\@tempb}{\@tempb:\@tempc}{\expandafter \expandafter \csname
  mn@eprint@\@tempb\endcsname \expandafter{\@tempc}}}

\bibitem[\protect\citeauthoryear{{Ahumada} et~al.,}{{Ahumada}
  et~al.}{2020}]{SDSS}
{Ahumada} R.,  et~al., 2020, \mn@doi [\apjs] {10.3847/1538-4365/ab929e}, \href
  {https://ui.adsabs.harvard.edu/abs/2020ApJS..249....3A} {249, 3}

\bibitem[\protect\citeauthoryear{{Arnaud}}{{Arnaud}}{1996}]{XSpec}
{Arnaud} K.~A.,  1996, in {Jacoby} G.~H.,  {Barnes} J.,  eds,  Astronomical
  Society of the Pacific Conference Series Vol. 101, Astronomical Data Analysis
  Software and Systems V. p.~17

\bibitem[\protect\citeauthoryear{{Baars}, {Genzel}, {Pauliny-Toth}  \&
  {Witzel}}{{Baars} et~al.}{1977}]{1977A&A....61...99B}
{Baars} J.~W.~M.,  {Genzel} R.,  {Pauliny-Toth} I.~I.~K.,   {Witzel} A.,  1977,
  \aap, \href {https://ui.adsabs.harvard.edu/abs/1977A&A....61...99B} {61, 99}

\bibitem[\protect\citeauthoryear{{Biava} et~al.,}{{Biava}
  et~al.}{2021}]{biava2021ultra}
{Biava} N.,  et~al., 2021, \mn@doi [\mnras] {10.1093/mnras/stab2840}, \href
  {https://ui.adsabs.harvard.edu/abs/2021MNRAS.508.3995B} {508, 3995}

\bibitem[\protect\citeauthoryear{{Briggs}}{{Briggs}}{1995}]{1995PhDT.......238B}
{Briggs} D.~S.,  1995, PhD thesis, New Mexico Institute of Mining and
  Technology

\bibitem[\protect\citeauthoryear{{Brunetti} \& {Jones}}{{Brunetti} \&
  {Jones}}{2014}]{2014IJMPD..2330007B}
{Brunetti} G.,  {Jones} T.~W.,  2014, \mn@doi [International Journal of Modern
  Physics D] {10.1142/S0218271814300079}, \href
  {https://ui.adsabs.harvard.edu/abs/2014IJMPD..2330007B} {23, 1430007}

\bibitem[\protect\citeauthoryear{{Brunetti}, {Setti}, {Feretti}  \&
  {Giovannini}}{{Brunetti} et~al.}{2001}]{Brunetti2001}
{Brunetti} G.,  {Setti} G.,  {Feretti} L.,   {Giovannini} G.,  2001, \mn@doi
  [\mnras] {10.1046/j.1365-8711.2001.03978.x}, \href
  {https://ui.adsabs.harvard.edu/abs/2001MNRAS.320..365B} {320, 365}

\bibitem[\protect\citeauthoryear{{Chambers} et~al.,}{{Chambers}
  et~al.}{2016}]{chambers2019panstarrs1}
{Chambers} K.~C.,  et~al., 2016, \mn@doi [arXiv e-prints]
  {10.48550/arXiv.1612.05560}, \href
  {https://ui.adsabs.harvard.edu/abs/2016arXiv161205560C} {p. arXiv:1612.05560}

\bibitem[\protect\citeauthoryear{{Clarke}, {Randall}, {Sarazin}, {Blanton}  \&
  {Giacintucci}}{{Clarke} et~al.}{2013}]{Clarke_2013}
{Clarke} T.~E.,  {Randall} S.~W.,  {Sarazin} C.~L.,  {Blanton} E.~L.,
  {Giacintucci} S.,  2013, \mn@doi [\apj] {10.1088/0004-637X/772/2/84}, \href
  {https://ui.adsabs.harvard.edu/abs/2013ApJ...772...84C} {772, 84}

\bibitem[\protect\citeauthoryear{{Cohen} \& {Clarke}}{{Cohen} \&
  {Clarke}}{2011}]{Cohen2011}
{Cohen} A.~S.,  {Clarke} T.~E.,  2011, \mn@doi [\aj]
  {10.1088/0004-6256/141/5/149}, \href
  {https://ui.adsabs.harvard.edu/abs/2011AJ....141..149C} {141, 149}

\bibitem[\protect\citeauthoryear{{Cohen}, {Lane}, {Cotton}, {Kassim}, {Lazio},
  {Perley}, {Condon}  \& {Erickson}}{{Cohen} et~al.}{2007}]{cohen2007}
{Cohen} A.~S.,  {Lane} W.~M.,  {Cotton} W.~D.,  {Kassim} N.~E.,  {Lazio}
  T.~J.~W.,  {Perley} R.~A.,  {Condon} J.~J.,   {Erickson} W.~C.,  2007,
  \mn@doi [\aj] {10.1086/520719}, \href
  {https://ui.adsabs.harvard.edu/abs/2007AJ....134.1245C} {134, 1245}

\bibitem[\protect\citeauthoryear{{Condon}, {Cotton}, {Greisen}, {Yin},
  {Perley}, {Taylor}  \& {Broderick}}{{Condon} et~al.}{1998}]{condon1998}
{Condon} J.~J.,  {Cotton} W.~D.,  {Greisen} E.~W.,  {Yin} Q.~F.,  {Perley}
  R.~A.,  {Taylor} G.~B.,   {Broderick} J.~J.,  1998, \mn@doi [\aj]
  {10.1086/300337}, \href
  {https://ui.adsabs.harvard.edu/abs/1998AJ....115.1693C} {115, 1693}

\bibitem[\protect\citeauthoryear{{Cornwell} \& {Perley}}{{Cornwell} \&
  {Perley}}{1992}]{1992A&A...261..353C}
{Cornwell} T.~J.,  {Perley} R.~A.,  1992, \aap, \href
  {https://ui.adsabs.harvard.edu/abs/1992A&A...261..353C} {261, 353}

\bibitem[\protect\citeauthoryear{{En{\ss}lin} \& {Br{\"u}ggen}}{{En{\ss}lin} \&
  {Br{\"u}ggen}}{2002}]{ensslin2002formation}
{En{\ss}lin} T.~A.,  {Br{\"u}ggen} M.,  2002, \mn@doi [\mnras]
  {10.1046/j.1365-8711.2002.05261.x}, \href
  {https://ui.adsabs.harvard.edu/abs/2002MNRAS.331.1011E} {331, 1011}

\bibitem[\protect\citeauthoryear{{En{\ss}lin} \& {Gopal-Krishna}}{{En{\ss}lin}
  \& {Gopal-Krishna}}{2001}]{Enblin2001}
{En{\ss}lin} T.~A.,  {Gopal-Krishna} 2001, \mn@doi [\aap]
  {10.1051/0004-6361:20000198}, \href
  {https://ui.adsabs.harvard.edu/abs/2001A&A...366...26E} {366, 26}

\bibitem[\protect\citeauthoryear{{Feretti}, {Giovannini}, {Govoni}  \&
  {Murgia}}{{Feretti} et~al.}{2012}]{Feretti2012}
{Feretti} L.,  {Giovannini} G.,  {Govoni} F.,   {Murgia} M.,  2012, \mn@doi
  [\aapr] {10.1007/s00159-012-0054-z}, \href
  {https://ui.adsabs.harvard.edu/abs/2012A&ARv..20...54F} {20, 54}

\bibitem[\protect\citeauthoryear{Foster, Smith  \& Brickhouse}{Foster
  et~al.}{2017}]{AtomDB}
Foster A.,  Smith R.,   Brickhouse N.,  2017, AIP Conference Proceedings, 1811

\bibitem[\protect\citeauthoryear{{Giacintucci} et~al.,}{{Giacintucci}
  et~al.}{2008}]{Giacintucci2008}
{Giacintucci} S.,  et~al., 2008, \mn@doi [\aap] {10.1051/0004-6361:200809459},
  \href {https://ui.adsabs.harvard.edu/abs/2008A&A...486..347G} {486, 347}

\bibitem[\protect\citeauthoryear{{Giacintucci}, {Kale}, {Wik}, {Venturi}  \&
  {Markevitch}}{{Giacintucci} et~al.}{2013}]{giacintucci2013discovery}
{Giacintucci} S.,  {Kale} R.,  {Wik} D.~R.,  {Venturi} T.,   {Markevitch} M.,
  2013, \mn@doi [\apj] {10.1088/0004-637X/766/1/18}, \href
  {https://ui.adsabs.harvard.edu/abs/2013ApJ...766...18G} {766, 18}

\bibitem[\protect\citeauthoryear{{Giacintucci}, {Markevitch}, {Cassano},
  {Venturi}, {Clarke}  \& {Brunetti}}{{Giacintucci}
  et~al.}{2017}]{2017ApJ...841...71G}
{Giacintucci} S.,  {Markevitch} M.,  {Cassano} R.,  {Venturi} T.,  {Clarke}
  T.~E.,   {Brunetti} G.,  2017, \mn@doi [\apj] {10.3847/1538-4357/aa7069},
  \href {https://ui.adsabs.harvard.edu/abs/2017ApJ...841...71G} {841, 71}

\bibitem[\protect\citeauthoryear{{Giacintucci}, {Markevitch}, {Cassano},
  {Venturi}, {Clarke}, {Kale}  \& {Cuciti}}{{Giacintucci}
  et~al.}{2019}]{2019ApJ...880...70G}
{Giacintucci} S.,  {Markevitch} M.,  {Cassano} R.,  {Venturi} T.,  {Clarke}
  T.~E.,  {Kale} R.,   {Cuciti} V.,  2019, \mn@doi [\apj]
  {10.3847/1538-4357/ab29f1}, \href
  {https://ui.adsabs.harvard.edu/abs/2019ApJ...880...70G} {880, 70}

\bibitem[\protect\citeauthoryear{{Govoni}, {Feretti}, {Giovannini},
  {B{\"o}hringer}, {Reiprich}  \& {Murgia}}{{Govoni}
  et~al.}{2001}]{govoni2001radio}
{Govoni} F.,  {Feretti} L.,  {Giovannini} G.,  {B{\"o}hringer} H.,  {Reiprich}
  T.~H.,   {Murgia} M.,  2001, \mn@doi [\aap] {10.1051/0004-6361:20011016},
  \href {https://ui.adsabs.harvard.edu/abs/2001A&A...376..803G} {376, 803}

\bibitem[\protect\citeauthoryear{{Greisen}}{{Greisen}}{2003}]{2003ASSL..285..109G}
{Greisen} E.~W.,  2003, in {Heck} A.,  ed.,  Astrophysics and Space Science
  Library Vol. 285, Information Handling in Astronomy - Historical Vistas.
  p.~109, \mn@doi{10.1007/0-306-48080-8_7}

\bibitem[\protect\citeauthoryear{{Grevesse} \& {Sauval}}{{Grevesse} \&
  {Sauval}}{1998}]{GrSa}
{Grevesse} N.,  {Sauval} A.~J.,  1998, \mn@doi [\ssr]
  {10.1023/A:1005161325181}, \href
  {https://ui.adsabs.harvard.edu/abs/1998SSRv...85..161G} {85, 161}

\bibitem[\protect\citeauthoryear{{HI4PI Collaboration} et~al.,}{{HI4PI
  Collaboration} et~al.}{2016}]{nH}
{HI4PI Collaboration} et~al., 2016, \mn@doi [\aap]
  {10.1051/0004-6361/201629178}, \href
  {https://ui.adsabs.harvard.edu/abs/2016A&A...594A.116H} {594, A116}

\bibitem[\protect\citeauthoryear{{Hodgson}, {Bartalucci}, {Johnston-Hollitt},
  {McKinley}, {Vazza}  \& {Wittor}}{{Hodgson} et~al.}{2021}]{hodgson2021ultra}
{Hodgson} T.,  {Bartalucci} I.,  {Johnston-Hollitt} M.,  {McKinley} B.,
  {Vazza} F.,   {Wittor} D.,  2021, \mn@doi [\apj] {10.3847/1538-4357/abe384},
  \href {https://ui.adsabs.harvard.edu/abs/2021ApJ...909..198H} {909, 198}

\bibitem[\protect\citeauthoryear{{Intema}, {Jagannathan}, {Mooley}  \&
  {Frail}}{{Intema} et~al.}{2017}]{intema2017}
{Intema} H.~T.,  {Jagannathan} P.,  {Mooley} K.~P.,   {Frail} D.~A.,  2017,
  \mn@doi [\aap] {10.1051/0004-6361/201628536}, \href
  {https://ui.adsabs.harvard.edu/abs/2017A&A...598A..78I} {598, A78}

\bibitem[\protect\citeauthoryear{{Ishibashi} \& {Courvoisier}}{{Ishibashi} \&
  {Courvoisier}}{2010}]{Ishibashi2010}
{Ishibashi} W.,  {Courvoisier} T.~J.~L.,  2010, \mn@doi [\aap]
  {10.1051/0004-6361/200913587}, \href
  {https://ui.adsabs.harvard.edu/abs/2010A&A...512A..58I} {512, A58}

\bibitem[\protect\citeauthoryear{{Kassim}, {Clarke}, {En{\ss}lin}, {Cohen}  \&
  {Neumann}}{{Kassim} et~al.}{2001}]{Kassim2001}
{Kassim} N.~E.,  {Clarke} T.~E.,  {En{\ss}lin} T.~A.,  {Cohen} A.~S.,
  {Neumann} D.~M.,  2001, \mn@doi [\apj] {10.1086/322381}, \href
  {https://ui.adsabs.harvard.edu/abs/2001ApJ...559..785K} {559, 785}

\bibitem[\protect\citeauthoryear{{Kempner}, {Blanton}, {Clarke}, {En{\ss}lin},
  {Johnston-Hollitt}  \& {Rudnick}}{{Kempner}
  et~al.}{2004}]{2004rcfg.proc..335K}
{Kempner} J.~C.,  {Blanton} E.~L.,  {Clarke} T.~E.,  {En{\ss}lin} T.~A.,
  {Johnston-Hollitt} M.,   {Rudnick} L.,  2004, in {Reiprich} T.,  {Kempner}
  J.,   {Soker} N.,  eds, The Riddle of Cooling Flows in Galaxies and Clusters
  of galaxies. p.~335 (\mn@eprint {arXiv} {astro-ph/0310263}),
  \mn@doi{10.48550/arXiv.astro-ph/0310263}

\bibitem[\protect\citeauthoryear{{Kuntz} \& {Snowden}}{{Kuntz} \&
  {Snowden}}{2008}]{ParticleBkg}
{Kuntz} K.~D.,  {Snowden} S.~L.,  2008, \mn@doi [\aap]
  {10.1051/0004-6361:20077912}, \href
  {https://ui.adsabs.harvard.edu/abs/2008A&A...478..575K} {478, 575}

\bibitem[\protect\citeauthoryear{{Lane}, {Cotton}, {van Velzen}, {Clarke},
  {Kassim}, {Helmboldt}, {Lazio}  \& {Cohen}}{{Lane} et~al.}{2014}]{lane2014}
{Lane} W.~M.,  {Cotton} W.~D.,  {van Velzen} S.,  {Clarke} T.~E.,  {Kassim}
  N.~E.,  {Helmboldt} J.~F.,  {Lazio} T.~J.~W.,   {Cohen} A.~S.,  2014, \mn@doi
  [\mnras] {10.1093/mnras/stu256}, \href
  {https://ui.adsabs.harvard.edu/abs/2014MNRAS.440..327L} {440, 327}

\bibitem[\protect\citeauthoryear{{Macario}, {Markevitch}, {Giacintucci},
  {Brunetti}, {Venturi}  \& {Murray}}{{Macario} et~al.}{2011}]{Macario2011}
{Macario} G.,  {Markevitch} M.,  {Giacintucci} S.,  {Brunetti} G.,  {Venturi}
  T.,   {Murray} S.~S.,  2011, \mn@doi [\apj] {10.1088/0004-637X/728/2/82},
  \href {https://ui.adsabs.harvard.edu/abs/2011ApJ...728...82M} {728, 82}

\bibitem[\protect\citeauthoryear{Mandal}{Mandal}{2020}]{mandal2020revealing}
Mandal S.,  2020, PhD thesis, Leiden University

\bibitem[\protect\citeauthoryear{{Markevitch}}{{Markevitch}}{2006}]{2006ESASP.604..723M}
{Markevitch} M.,  2006, in {Wilson} A.,  ed.,  ESA Special Publication Vol.
  604, The X-ray Universe 2005. p.~723 (\mn@eprint {arXiv} {astro-ph/0511345}),
  \mn@doi{10.48550/arXiv.astro-ph/0511345}

\bibitem[\protect\citeauthoryear{{Markevitch}, {Sarazin}  \&
  {Vikhlinin}}{{Markevitch} et~al.}{1999}]{Markevitch1999}
{Markevitch} M.,  {Sarazin} C.~L.,   {Vikhlinin} A.,  1999, \mn@doi [\apj]
  {10.1086/307598}, \href
  {https://ui.adsabs.harvard.edu/abs/1999ApJ...521..526M} {521, 526}

\bibitem[\protect\citeauthoryear{Mernier, Werner, Bagchi, Gendron-Marsolais,
  Guainazzi, Richard-Laferri{\`e}re, Shimwell  \& Simionescu}{Mernier
  et~al.}{2023}]{mernier2023discovery}
Mernier F.,  Werner N.,  Bagchi J.,  Gendron-Marsolais M.,  Guainazzi M.,
  Richard-Laferri{\`e}re A.,  Shimwell T.,   Simionescu A.,  2023, \mn@doi
  [MNRAS] {10.1093/mnras/stad2093}, 524, 4939

\bibitem[\protect\citeauthoryear{{Milosavljevi{\'c}}, {Koda}, {Nagai}, {Nakar}
  \& {Shapiro}}{{Milosavljevi{\'c}} et~al.}{2007}]{milosavljevic2007cluster}
{Milosavljevi{\'c}} M.,  {Koda} J.,  {Nagai} D.,  {Nakar} E.,   {Shapiro}
  P.~R.,  2007, \mn@doi [\apjl] {10.1086/518960}, \href
  {https://ui.adsabs.harvard.edu/abs/2007ApJ...661L.131M} {661, L131}

\bibitem[\protect\citeauthoryear{{Mohr}, {Mathiesen}  \& {Evrard}}{{Mohr}
  et~al.}{1999}]{Mohr1999}
{Mohr} J.~J.,  {Mathiesen} B.,   {Evrard} A.~E.,  1999, \mn@doi [\apj]
  {10.1086/307227}, \href
  {https://ui.adsabs.harvard.edu/abs/1999ApJ...517..627M} {517, 627}

\bibitem[\protect\citeauthoryear{Pacholczyk}{Pacholczyk}{1977}]{Pacholczyk1977}
Pacholczyk A.,  1977, Radio Galaxies: Radiation Transfer, Dynamics, Stability
  and Evolution of a Synchrotron Plasmon.
Elsevier Ltd

\bibitem[\protect\citeauthoryear{Perley}{Perley}{1989}]{1989ASPC....6.....P}
Perley R.,  ed. 1989, {Synthesis imaging in radio astronomy : a collection of
  lectures from the third NRAO synthesis imaging summer school}  Astronomical
  Society of the Pacific Conference Series Vol. 6

\bibitem[\protect\citeauthoryear{Perley \& Taylor}{Perley \&
  Taylor}{1999}]{perley1999vla}
Perley R.,  Taylor G.,  1999, Technical report, VLA Calibrator Manual.
Tech. rep., NRAO

\bibitem[\protect\citeauthoryear{{Petrosian}}{{Petrosian}}{2001}]{Petrosian2001}
{Petrosian} V.,  2001, \mn@doi [\apj] {10.1086/321557}, \href
  {https://ui.adsabs.harvard.edu/abs/2001ApJ...557..560P} {557, 560}

\bibitem[\protect\citeauthoryear{Randall et~al.,}{Randall
  et~al.}{2016}]{Randall2016}
Randall S.,  et~al., 2016, ApJ, 823, A94

\bibitem[\protect\citeauthoryear{{Richard-Laferri{\`e}re}
  et~al.,}{{Richard-Laferri{\`e}re} et~al.}{2020}]{2020MNRAS.499.2934R}
{Richard-Laferri{\`e}re} A.,  et~al., 2020, \mn@doi [\mnras]
  {10.1093/mnras/staa2877}, \href
  {https://ui.adsabs.harvard.edu/abs/2020MNRAS.499.2934R} {499, 2934}

\bibitem[\protect\citeauthoryear{Riseley et~al.,}{Riseley
  et~al.}{2022}]{10.1093/mnras/stac672}
Riseley C.~J.,  et~al., 2022, \mn@doi [Monthly Notices of the Royal
  Astronomical Society] {10.1093/mnras/stac672}, 512, 4210

\bibitem[\protect\citeauthoryear{{Sabol} \& {Snowden}}{{Sabol} \&
  {Snowden}}{2019}]{RASStool}
{Sabol} E.~J.,  {Snowden} S.~L.,  2019, {sxrbg: ROSAT X-Ray Background Tool},
  Astrophysics Source Code Library, record ascl:1904.001 (\mn@eprint {ascl}
  {1904.001})

\bibitem[\protect\citeauthoryear{{Salunkhe}, {Paul}, {Krishna}, {Sonkamble}  \&
  {Bhagat}}{{Salunkhe} et~al.}{2022}]{salunkhe2022deciphering}
{Salunkhe} S.,  {Paul} S.,  {Krishna} G.,  {Sonkamble} S.,   {Bhagat} S.,
  2022, \mn@doi [\aap] {10.1051/0004-6361/202243438}, \href
  {https://ui.adsabs.harvard.edu/abs/2022A&A...664A.186S} {664, A186}

\bibitem[\protect\citeauthoryear{{Sanders}}{{Sanders}}{2006}]{Sanders2006}
{Sanders} J.~S.,  2006, \mn@doi [\mnras] {10.1111/j.1365-2966.2006.10716.x},
  \href {https://ui.adsabs.harvard.edu/abs/2006MNRAS.371..829S} {371, 829}

\bibitem[\protect\citeauthoryear{{Sanders} et~al.,}{{Sanders}
  et~al.}{2016a}]{GGM1}
{Sanders} J.~S.,  et~al., 2016a, \mn@doi [\mnras] {10.1093/mnras/stv2972},
  \href {https://ui.adsabs.harvard.edu/abs/2016MNRAS.457...82S} {457, 82}

\bibitem[\protect\citeauthoryear{{Sanders}, {Fabian}, {Russell}, {Walker}  \&
  {Blundell}}{{Sanders} et~al.}{2016b}]{GGM2}
{Sanders} J.~S.,  {Fabian} A.~C.,  {Russell} H.~R.,  {Walker} S.~A.,
  {Blundell} K.~M.,  2016b, \mn@doi [\mnras] {10.1093/mnras/stw1119}, \href
  {https://ui.adsabs.harvard.edu/abs/2016MNRAS.460.1898S} {460, 1898}

\bibitem[\protect\citeauthoryear{{Santra} et~al.,}{{Santra}
  et~al.}{2023}]{santra2023deep}
{Santra} R.,  et~al., 2023, \mn@doi [arXiv e-prints]
  {10.48550/arXiv.2311.09717}, \href
  {https://ui.adsabs.harvard.edu/abs/2023arXiv231109717S} {p. arXiv:2311.09717}

\bibitem[\protect\citeauthoryear{{Sarazin}}{{Sarazin}}{1999}]{Sarazin1999}
{Sarazin} C.~L.,  1999, \mn@doi [\apj] {10.1086/307501}, \href
  {https://ui.adsabs.harvard.edu/abs/1999ApJ...520..529S} {520, 529}

\bibitem[\protect\citeauthoryear{{Sargent}}{{Sargent}}{1972}]{sargent1972cluster}
{Sargent} W. L.~W.,  1972, \mn@doi [\apj] {10.1086/151659}, \href
  {https://ui.adsabs.harvard.edu/abs/1972ApJ...176..581S} {176, 581}

\bibitem[\protect\citeauthoryear{{Shimwell}, {Brown}, {Feain}, {Feretti},
  {Gaensler}  \& {Lage}}{{Shimwell} et~al.}{2014}]{Shimwell2014}
{Shimwell} T.~W.,  {Brown} S.,  {Feain} I.~J.,  {Feretti} L.,  {Gaensler}
  B.~M.,   {Lage} C.,  2014, \mn@doi [\mnras] {10.1093/mnras/stu467}, \href
  {https://ui.adsabs.harvard.edu/abs/2014MNRAS.440.2901S} {440, 2901}

\bibitem[\protect\citeauthoryear{{Shimwell} et~al.,}{{Shimwell}
  et~al.}{2019}]{shimwell2019}
{Shimwell} T.~W.,  et~al., 2019, \mn@doi [\aap] {10.1051/0004-6361/201833559},
  \href {https://ui.adsabs.harvard.edu/abs/2019A&A...622A...1S} {622, A1}

\bibitem[\protect\citeauthoryear{{Shimwell} et~al.,}{{Shimwell}
  et~al.}{2022}]{shimwell2022}
{Shimwell} T.~W.,  et~al., 2022, \mn@doi [\aap] {10.1051/0004-6361/202142484},
  \href {https://ui.adsabs.harvard.edu/abs/2022A&A...659A...1S} {659, A1}

\bibitem[\protect\citeauthoryear{{Slee}, {Roy}, {Murgia}, {Andernach}  \&
  {Ehle}}{{Slee} et~al.}{2001}]{Slee2001}
{Slee} O.~B.,  {Roy} A.~L.,  {Murgia} M.,  {Andernach} H.,   {Ehle} M.,  2001,
  \mn@doi [\aj] {10.1086/322105}, \href
  {https://ui.adsabs.harvard.edu/abs/2001AJ....122.1172S} {122, 1172}

\bibitem[\protect\citeauthoryear{Snowden \& Kuntz}{Snowden \&
  Kuntz}{2014}]{cookbook}
Snowden S.,  Kuntz K.,  2014, XMM-Newton ESAS Cookbook

\bibitem[\protect\citeauthoryear{{Snowden} et~al.,}{{Snowden}
  et~al.}{1997}]{RASS}
{Snowden} S.~L.,  et~al., 1997, \mn@doi [\apj] {10.1086/304399}, \href
  {https://ui.adsabs.harvard.edu/abs/1997ApJ...485..125S} {485, 125}

\bibitem[\protect\citeauthoryear{{Struble} \& {Rood}}{{Struble} \&
  {Rood}}{1991}]{Struble1991}
{Struble} M.~F.,  {Rood} H.~J.,  1991, \mn@doi [\apjs] {10.1086/191608}, \href
  {https://ui.adsabs.harvard.edu/abs/1991ApJS...77..363S} {77, 363}

\bibitem[\protect\citeauthoryear{{Vikhlinin} et~al.,}{{Vikhlinin}
  et~al.}{2009}]{r500}
{Vikhlinin} A.,  et~al., 2009, \mn@doi [\apj] {10.1088/0004-637X/692/2/1033},
  \href {https://ui.adsabs.harvard.edu/abs/2009ApJ...692.1033V} {692, 1033}

\bibitem[\protect\citeauthoryear{{Wik} et~al.,}{{Wik} et~al.}{2014}]{Wik2014}
{Wik} D.~R.,  et~al., 2014, \mn@doi [\apj] {10.1088/0004-637X/792/1/48}, \href
  {https://ui.adsabs.harvard.edu/abs/2014ApJ...792...48W} {792, 48}

\bibitem[\protect\citeauthoryear{{ZuHone}}{{ZuHone}}{2011}]{zuhone2011parameter}
{ZuHone} J.~A.,  2011, \mn@doi [\apj] {10.1088/0004-637X/728/1/54}, \href
  {https://ui.adsabs.harvard.edu/abs/2011ApJ...728...54Z} {728, 54}

\bibitem[\protect\citeauthoryear{{de Gasperin}, {van Weeren}, {Br{\"u}ggen},
  {Vazza}, {Bonafede}  \& {Intema}}{{de Gasperin} et~al.}{2014}]{de2014new}
{de Gasperin} F.,  {van Weeren} R.~J.,  {Br{\"u}ggen} M.,  {Vazza} F.,
  {Bonafede} A.,   {Intema} H.~T.,  2014, \mn@doi [\mnras]
  {10.1093/mnras/stu1658}, \href
  {https://ui.adsabs.harvard.edu/abs/2014MNRAS.444.3130D} {444, 3130}

\bibitem[\protect\citeauthoryear{{de Gasperin}, {Ogrean}, {van Weeren},
  {Dawson}, {Br{\"u}ggen}, {Bonafede}  \& {Simionescu}}{{de Gasperin}
  et~al.}{2015}]{Gasperin2015}
{de Gasperin} F.,  {Ogrean} G.~A.,  {van Weeren} R.~J.,  {Dawson} W.~A.,
  {Br{\"u}ggen} M.,  {Bonafede} A.,   {Simionescu} A.,  2015, \mn@doi [\mnras]
  {10.1093/mnras/stv129}, \href
  {https://ui.adsabs.harvard.edu/abs/2015MNRAS.448.2197D} {448, 2197}

\bibitem[\protect\citeauthoryear{{van Weeren}, {R{\"o}ttgering}  \&
  {Br{\"u}ggen}}{{van Weeren} et~al.}{2011a}]{2011A&A...527A.114V}
{van Weeren} R.~J.,  {R{\"o}ttgering} H.~J.~A.,   {Br{\"u}ggen} M.,  2011a,
  \mn@doi [\aap] {10.1051/0004-6361/201015991}, \href
  {https://ui.adsabs.harvard.edu/abs/2011A&A...527A.114V} {527, A114}

\bibitem[\protect\citeauthoryear{{van Weeren}, {Hoeft}, {R{\"o}ttgering},
  {Br{\"u}ggen}, {Intema}  \& {van Velzen}}{{van Weeren}
  et~al.}{2011b}]{2011A&A...528A..38V}
{van Weeren} R.~J.,  {Hoeft} M.,  {R{\"o}ttgering} H.~J.~A.,  {Br{\"u}ggen} M.,
   {Intema} H.~T.,   {van Velzen} S.,  2011b, \mn@doi [\aap]
  {10.1051/0004-6361/201016185}, \href
  {https://ui.adsabs.harvard.edu/abs/2011A&A...528A..38V} {528, A38}

\bibitem[\protect\citeauthoryear{{van Weeren}, {de Gasperin}, {Akamatsu},
  {Br{\"u}ggen}, {Feretti}, {Kang}, {Stroe}  \& {Zandanel}}{{van Weeren}
  et~al.}{2019}]{van2019diffuse}
{van Weeren} R.~J.,  {de Gasperin} F.,  {Akamatsu} H.,  {Br{\"u}ggen} M.,
  {Feretti} L.,  {Kang} H.,  {Stroe} A.,   {Zandanel} F.,  2019, \mn@doi [\ssr]
  {10.1007/s11214-019-0584-z}, \href
  {https://ui.adsabs.harvard.edu/abs/2019SSRv..215...16V} {215, 16}

\makeatother
\end{thebibliography}

\bsp	% typesetting comment
\label{lastpage}
\end{document}